\documentclass{article}

\usepackage{amsmath,amssymb,dsfont} % define this before the line numbering.
\usepackage{graphicx}
\usepackage{balance}
\graphicspath{{figures}{tikzexternal}}
\usepackage{pgfplots}
\pgfplotsset{compat=newest}
\usetikzlibrary{spy,positioning,shapes,calc,%
                intersections,through,backgrounds,%
                decorations.text,plotmarks,external}
\usepgfplotslibrary{units}
\tikzset{external/system call={lualatex \tikzexternalcheckshellescape -halt-on-error -interaction=batchmode -jobname "\image" "\texsource"}}
\newlength\figureheight
\newlength\figurewidth
\newcommand{\titlefortikz}{Title}
\newcommand{\xlabelfortikz}{Time}
\newcommand{\ylabelfortikz}{Range-difference}
\usepackage{color}
\usepackage{cite,times, epsfig, url, array, algorithmic}
\usepackage{epstopdf}
\usepackage{xspace}
\usepackage{mathtools}
\usepackage{subcaption}
\usepackage{multirow}
\usepackage{enumitem}
\usepackage{tikz}
\usetikzlibrary{matrix,positioning,decorations.pathreplacing}

\DeclareRobustCommand\onedot{\futurelet\@let@token\@onedot}
\def\onedot{.\xspace}

\newcommand{\fatR}{\mathbb{R}}

\def\rank{\operatorname{rank}}

\def\ltwo{{$l^2$}}
\newcommand{\rrr}{\mathbf{r}}
\newcommand{\sss}{\mathbf{s}}

\providecommand{\norm}[1]{\lVert#1\rVert}

\makeatletter

\newcommand{\Rmnum}[1]{\expandafter\@slowromancap\romannumeral #1@}
\makeatother

\begin{document}

\title{An Automatic System for Acoustic Microphone Geometry Calibration based on Minimal Solvers}

\author{Simayijiang~Zhayida,
        Simon~Segerblom~Rex,
        Yubin~Kuang,\\
        Fredrik~Andersson,
        and~Kalle~{\AA}str{\"o}m\thanks{The work was supported by the strategic research projects ELLIIT and eSSENCE and the Swedish Foundation for Strategic Research (grant no. RIT15-0038), the Swedish Foundation for International Cooperation in Research and Higher Education, the Craaford Foundation and the Swedish Research Council.}\\
Centre for Mathematical Sciences, Lund University, \\
Box118, SE-22100, LUND, SWEDEN
%\thanks{Corresponding author affiliation: Centre for Mathematical Sciences, Lund Institute of Technology/Lund University, Box118, SE-22100, LUND, SWEDEN.
%Tel.: +46 46 22 285 37, Fax: +46 46 22 240 10.}% <-this % stops a space
%\thanks{ Email: \{zhayida, kalle\}@maths.lth.se}% <-this % stops a space
%\thanks{This work is an extension of \cite{kuang-astrom-2espc2-13,zhayida-etc-eusipco2014}, published at the conferences EUSIPCO 2013 and 2014.}
}
\maketitle

\begin{abstract}
In this paper, robust detection, tracking and geometry estimation methods are developed and combined into a system for estimating time-difference estimates, microphone localization and sound source movement. 
No assumptions on the 3D locations of the microphones and sound sources are made. The system is capable of tracking continuously moving sound sources in an reverberant environment. The multi-path components are explicitly tracked and used in the geometry estimation parts. The system is based on matching between pairs of channels using GCC-PHAT. Instead of taking a single maximum at each time instant from each such pair, we select the four strongest local maxima. This produce a set of hypothesis to work with in the subsequent steps, where consistency constraints between the channels and time-continuity constraints are exploited. In the paper it demonstrated how such detections can be used to estimate microphone positions, sound source movement and room geometry. The methods are tested and verified using real data from several reverberant environments. The evaluation demonstrated accuracy in the order of few millimeters. 
\end{abstract}

\section{Introduction}
\label{sec:intro}

Many of our everyday tools, such as smartphones, laptops, and tablet pc's, are equipped with microphones. Given the location of each one of these microphones, it is possible to use them as ad-hoc acoustic sensor network. Such sensor networks can be used for many interesting applications. One application is to improve the sound quality using so called beam-forming. Another application is so called speaker diarization, i.e.\ to determine who spoke when. Networks can also be used for localization and mapping. If the microphone positions are unknown or only known to a certain accuracy, the results are inferior as is shown in \cite{plinge2016acoustic}. 

Since the problem of acoustic microphone geometry calibration is essential to applications such as speaker localization, speaker tracking, indoor navigation, beamforming and speaker diarization, the problem has recently become a very active field of research, see \cite{plinge2016acoustic}, which gives an overview of the research field and their applications.

It is sometimes useful to think of the problem as having a signal processing part and a geometry estimation part. In the signal processing part one tries to obtain measurements from the raw audio measurements. These measurements will have both small measurement errors, but also gross errors or outliers. In the geometry estimation part one tries to estimate microphone and sound source positions from the noisy measurements extracted from the audio signals. There is also a correspondence problem. One has to track these sound events over channels and over time. This correspondence problem could be solved in the signal processing part, in the geometry estimation part or a combination of the two. 

If the microphone positions are not known we obtain a chicken-and-egg kind of problem. Sound source localization is relatively easy if the microphone positions are known, and similarly microphone positions are relatively easy to solve if the sound source positions are known. But getting a sense of the data when the position of both microphones and sound sources are unknown is a more challenging problem. Furthermore if the sounds emitted are of unknown character and if the environment is reverberant, it is difficult to match the sounds from one sound source to the other. 

The signal processing part is substantially easier if the sounds emitted are short in duration and if they have been specifically designed to be easily detected. Popular examples here are chirps or claps, cf.\ \cite{kuang-astrom-2espc2-13}. 
%This makes the detection of the claps and chirps easier.

In this paper we present new methods for simultaneous signal processing and geometry estimation such as to obtain estimates of microphone positions, sound source positions and major reverberant planar structures, without requiring that the sound sources has a particular structure, for instance that they are of short duration. 

\paragraph{Literature survey}

% TOA
%There are various existing location
%estimation approaches, such as received signal strength intensity (RSSI), angle of
%arrival (AOA), time of arrival (TOA) and time difference of
%arrival (TDOA). Among these techniques, the ranged-based
%schemes, TOA and TDOA, are proved to have a very good
%accuracy [1288] due to the high time resolution (large bandwidth)
%of the signals.
%For the range-based schemes, the parameter extraction step is
%to gain ranges between the sensor nodes, so it can be called
%range estimation. The data fusion is the location estimation
%step.
% TOA

% Speaker tracking
The literature within sound source localization using known microphone positions is very large, see e.g.\ 
 \cite{Brandstein1997,Cirillo2008,Cobos2011,Hoang2007}, and the references therein.
 
If the sounds were generated by a speaker and if the speakers and the microphones are connected to the same sound card it would possible to measure the Time-of-Arrival (TOA) directly, e.g.\ by using generalized cross correlation techniques between the emitted and received sound signals, \cite{knapp76}. Examples of such experiments are given e.g.\ in \cite{Burgess-diffdim-2015}. Similarily there are applications, e.g.\ using UWB or so called fine time measurements, also using a radio technique, which allows for time of arrival measurements, \cite{batstone_icc_2016}. In these appliations the signal processing part is implemented in hardware and TOA measurements are produced. For such applications, only the geometric estimation part needs to be solved.

Assuming that Time-of-Arrival (TOA) measurements are given, the problem of determining both microphone and sender positions is still challenging. Early attempts at solving this chicken-and-egg type of problem often resorted in adding extra assumpions on the location of some of the sensors or assumptions on one of the sound sources being a the same position as one of the microphones, cf.\ \cite{birchfield-subramanya-sap,niculescu-nath,raykar-etal, crocco-delbue-etal-icassp2012,chen-etal-sp-2002,pertila2013passive}. 

Once a reasonable initial estimate of the positions are known, the problem can be formulated as a robust non-linear least squares problem, for which iterative methods exist, cf.\  \cite{biswas-thrun-iros04,wendeberg-etal-2011}. However, such methods depend on initialization and can get stuck in local minima.  For a general graph structure, one can relax the TOA-based calibration problem as a semi-definite program \cite{biswas2006semidefinite}. 

For TOA measurements there is also considerable knowledge on minimal solvers. Minimal solvers for linearly constrained motion is given in \cite{kuang-ask-etal-icpram-12}. The problem is well understood for the case of receivers or senders spanning sub-spaces of different dimension, \cite{Burgess-diffdim-2015}. When both receivers and senders span the same linear space, the minimal problems are 3 receivers and 3 senders in a plane, \cite{stewenius-phd-2005}. For the three-dimensional case the minimal problems are either 4 receivers and 6 senders (or symmetrically 6 receivers and 4 senders). The case of 5 receivers and 5 senders is close to minimal. Efficient algorithms for solving these problems are given in \cite{kuang-burgess-etal-icassp13}.
In \cite{Thrun05c} and refined in \cite{kuang-ask-etal-icpram-12} a far field approximation 
was utilized to initialize both TOA and TDOA problems. 
In the far-field approximation, it is assumed that the distances between the speakers and receivers is considerable larger than between receivers.

Initialization of TDOA networks is studied in
\cite{pollefeys-nister-icassp-08}, where a solution to
non-minimal case of 9 receivers and 4 speakers in 3D was presented. 
   
In \cite{valente2010self} a system is presented for calibrating two microphone arrays relative to each other. Here it is assumed that each array has been pre-calibrated first. The authors also assume that the environment in non-reverberant. This makes the signal processing easier. The authors calculate 3D positions of sound sources by maximizing the steered response power. This is essentially a 3D search, which is made by searching 3D space using a regular grid. Once a number of 3D matches are obtained the transformation between the two calibrated microphone arrays can be made. 

In \cite{schmalenstroeer2011unsupervised} it is assumed that there are several subgroups of pre-calibrated linear arrays, that all arrays and all sound sources are in a plane. Using beamforming one can determine the angle to the sound sources from each linear array. This gives rise to a 2D structure from motion problem, similar to that of 1D retinal computer vision, \cite{faugeras-quan-etal-5eccv-98,astrom-oskarsson-jmiv-00,astrom-kahl-jmiv-03}. The authors propose a method of several random initializations of the unknown parameters followed by non-linear least squares optimization.

In \cite{mccowan2008microphone}, the authors introduce the idea of using the diffuse noise field for estimating the distance between two microphones is introduced. Given a diffuse noise field the coherence function is real valued ans sinc-shaped with a parameter that depends on the distance between the two microphones. For this special case it is possible to estimate the parameter for every microphone pair and then use multi-dimensional scaling,\cite{young-householder-1938}, to obtain the microphone geometry. This was later used in \cite{hennecke2009hierarchical} together with RAndom SAmple Consensus (RANSAC) to obtain a more robust method. 

%\cite{jiang-etal-icassp13}
%\begin{figure}
%\centering
%\def\svgwidth{5cm}
%\includegraphics[width= 0.85 \columnwidth]{Illustration.pdf}
%\caption{The paper presents a automatic system for microphone self-localization based on ambient sound. In the experiments we use several moving sound sources as depicted in the figure. The aim is to construct systems that takes as input a sound recording from each of $m$ microphones.  Here we do not assume specialized sound sources, known sound source patterns. Neither do we put any restrictions on how the sound source moves.}
%\label{fig::illustration}
%%\vspace{-6mm}
%\end{figure}

\paragraph{This contribution}

In this paper, which is a combination and extension of two previous conference 
publications \cite{kuang-astrom-2espc2-13,zhayida-etc-eusipco2014}, 
we focus on a system approach. For the system we assume that
the microphones can be in unknown general 3D positions and that the microphones are synchronized. We do not
put any constraints on the sound sources. They can also be
in general 3D positions and do not have to fulfill the far-field
criteria. There can be multiple possible moving sound
sources. The input to the system is a sound recording with $m$ channels (one from each microphone). %see Figure~\ref{fig::illustration}. 
These are first processed to find the time-difference vectors. The estimated time-difference vectors typically contain noise, missing data and possible outliers. We then follow a stratified approach, where we first estimate offsets using a robust method. This is followed by a robust method for finding the 3D positions of all the microphones and the 3D positions of all sound sources. Next is refinement of matching matrix by finding more inliers from raw matches and also refinement of correlation pattern using sub-pixel methods, cf.\ \cite{astrom-heyden-aap-99}. 
%Block diagram of procedure can be seen on  Figure~\ref{fig::block_diagram}.

The main contributions of this paper are:
\begin{itemize}
\item New efficient algorithms for solving for offsets given tdoa data for 5 minimal problems.
\item New robust algorithms for solving the geometric TDOA problem given noisy data with outliers and missing data. 
\item New matching and tracking algorithms for obtaining TDOA data from acoustic signals.
\item New robust algorithms for determining reverbant planes.
\item Combination of these parts into a system.
\item Development of a database of test cases with ground truth to test the system.
\end{itemize}

%We emphasize the need for further research within this area of systems for automatic calibration. Code examples are given at github project \url{http://github.com/kalleastrom/StructureFromSound} and example sound recordings are available for download at \url{http://www.maths.lu.se/staff/kalleastrom/downloads/structure-from-sound-data}. We hope that such code and data can be used by other researchers to empirically test algorithms for feature detection and matching as well as for estimating offsets and geometry of the microphone-speaker setup. We believe that there are several interesting research questions and that there is substantial room for improvement both on the different parts of the system as well as on how to combine such parts into a system.

\section{System Design}
\label{sec:models}

\begin{figure}
\centering
%\begin{center}
  \setlength\figureheight{0.48\linewidth}
  \setlength\figurewidth{0.85\linewidth}
  \tikzsetnextfilename{bassh_still}
  \input{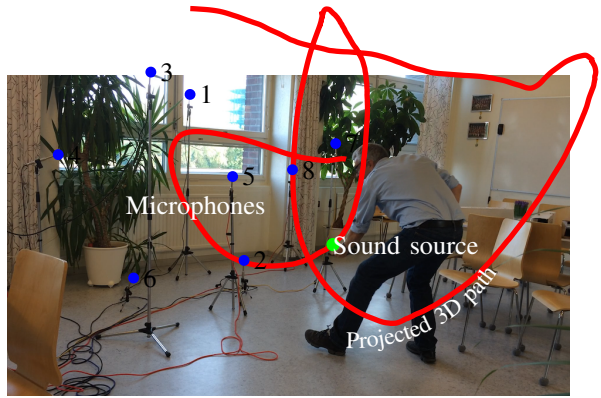}
  %\end{center}
  \caption{The paper presents a complete system for analysing sound recordings from $m$ microphones. The figure illustrates one of the real experiments. In this experiment we used one moving sound source, shown as a red path. In the experiment there are reverberations in the floor and the window, and numerous weaker reverberations caused by chairs, tables, humans etc. The system calculates microphone positions, room geometry and sound source motion.}
  \label{fig:illustration}
  %\vspace{-3mm}
\end{figure}

In this paper we present new tools and a first version of a system that can estimate (i) microphone positions, (ii) position and/or motion of one or several sound sources, and (iii) room geometry. We here assume that the number of simultaneous sound sources are few. In the experiments, we have mostly used a single moving sound source. We assume that the sound sources have relatively small spatial size and that the sounds are relatively omni-directional. We also assume that the microphones are relatively omnidirectional. Furthermore we assume that major reverberant surfaces are few and planar, modelled as planes $\pi_I$. The system is tested in a real environment, e.g.\ as illustrated in Figure~\ref{fig:illustration}. For this reason, it important that the system is robust to additional reverberations caused by other objects (chairs, people, etc) and robust to background sound sources, e.g. shoes and muffled talk. 

Contrary to many previous papers, we do not assume the simpler case of microphones being placed close to each other, so that a far-field approximation can be used. Also we do not assume that the microphones are placed close to each other so that one may assume that they are within a wavelength of each other. 

%The system is based on several components.
%First the detection, matching and tracking of time-difference data. 

The input to the system consists of $M$ sound recordings, 
$$y_i(\tau T_s), \tau = 1, \ldots, T, i = 1, \ldots, M, $$
where $i$ is used as index for the recordings and $\tau$ denotes the index for the sampling point and $T_s$ is the sampling period. % (time difference between two consecutive samples).
The sound sources are assumed to be spatially small and omnidirectional, so that they can be modelled as points moving in space. Each such sound event have an onset time $t_0$ and offset time $t_1$. During this interval $t \in [ t_0, t_1]$ the sound source moves in space according to $\mathbf{s} :  [ t_0, t_1] \rightarrow \fatR^3$. during this interval the emitted sound is given by $x(t)$. 
According to the model, at each time, the measured signals
\begin{equation}
y_i(t) \approx \sum_k \alpha_{i,k}(t) x(t - \tau_{i,k}(t)),
\label{eq:sound_model}
\end{equation}
is approximately a sum of a few scaled and time-delayed versions of the sound $x(t)$, emitted by the sound source plus noise, e.g. by stray sounds and reverberations not modelled by the main reflective planes. The time-delays $\tau_{i,k}$ are themselves time dependent. However, during the analysis of short time-intervals, they can be considered to be almost constant. Here it is assumed that the speed of the sound sources is much less than the speed of sound. We use an index $k$ to denote the index of the reflections, where $k=1$ is used to denote the direct path and higher indices are used for single and then multiple reflections in the planar surfaces.% , see Figure~\ref{fig:mirror}.

%The microphones are at unknown positions $(\bf \mathbf{r}_1, \ldots, \mathbf{s}_m)$, see Figure~\ref{fig::system}.
%Multipath components for reflections in dominant planar structures such as floors, ceilings, tables and walls can be modelled as microphones in mirrored positions, see Figure~\ref{fig::system}. These virtual (or multipath) microphone positions,
%$\mathbf{r}^k_i$ in Figure~\ref{fig::system},

Let  $\mathbf{r}_i$, $ i = 1, \dots, m$ be the spatial coordinates 
of the $m$ receivers. For the case of acoustic mirrored microphone we will use the notation $\mathbf{r}_{i,k}$ for the mirrored version of microphone $i$ corresponding to multi-path component $k$.
Let $\mathbf{s}$, be the spatial coordinate vector of a position at which a sound was emitted for a sound event at a particular time. 
For measured time of arrival $\tau_{ik}$ from sound transmission at a position $\mathbf{s}$ with unknown emission time $t_{emission}$ to receiver at a position $\mathbf{r}_{i,k}$, we have that $v \tau_{i,k}  = {\norm{\mathbf{r}_{i,k} - \mathbf{s}}}_2$, where $v$ is the speed of measured signals and ${\norm{\cdot}}_2$ is the \ltwo-norm. The speed $v$ is assumed to be known and constant. 

If the sounds were generated by a speaker and if the speakers and the microphones are connected to the same sound card it would possible to measure these time of arrivals directly. This is possible since then both functions $y$ and $x$ in Equation~\ref{eq:sound_model} would be known. So by generalized cross correlation techniques it is possible to determine $\tau_{i,k}$ from $x$ and $y$, \cite{knapp76}. In this paper, however, we consider the problem of listening to ambient sound. We assume that the microphones (the receivers) are connected to the same sound card, or that they are synchronized. This assumption could in fact be relaxed using calibration techniques to synchronize the data afterwards, e.g.\ using techniques developed in \cite{burgess2013minimal}. This extension would allow the use of arbitrary non-synchronized recording devices. For the rest of the paper, however, we assume that the receivers are synchronized but that the sound sources are unknown and thus unsynchronized. Thus, we are forced to use time difference measurements. 

Two channels can be matched using different techniques, e.g.\ cross-correlation or generalized cross-correlation. Generalized cross-correlation method can weaken the impact of noise on the delay estimation accuracy \cite{knapp76,liu2012research}. Here we opted to use GCC-PHAT for the estimation of time-differences. 
The matching score between sounds $i_1$ and $i_2$ typically have peaks for relative time delays $\tau_{i_2,k_2}-\tau_{i_1,k_1}$, for different combinations of multi-path components $k_1$ and $k_2$ for the two microphones in question, see (Equation~\ref{eq:sound_model}). Our experience is that the difference for the direct path, i.e.\ $\tau_{i_2,1}-\tau_{i_1,1}$ is more likely have a high matching score. However, due to noise it is often the case that this direct path does not yield the strongest peak and sometimes the direct path does not yield a peak at all. In this paper we will usually scale the time-differences with speed so as to obtain range-differences. The connection between the range difference and the geometric positions of the sound source and the receivers are then 
\begin{equation}
w_{i_1,k_1,i_2,k_2} = v(\tau_{i_2,k_2}-\tau_{i_1,k_1}) = || \sss - \rrr_{i_2,k_2} || - || \sss - \rrr_{i_1,k_1} || . 
\label{eq:uikik}
\end{equation}

In principle one could study correlation techniques that study three or more channels directly. But the search space and computational complexity then becomes a problem. For two channels each time-difference has one degree of freedom. This one degree of freedom is determined efficiently using the fast Fourier transform. For $m$ channels each multi-path component has $m-1$ degrees of freedom. Our hypothesis is that using pairwise channels gives the best trade-off between computational complexity and performance. However, in the future, it would be interesting to further investigate techniques that explicitly study multi-channel correlation techniques.  
%\begin{figure}[h]
%	\centering
%	\def\svgwidth{10cm}
%	\includegraphics[height=6cm,width= 8cm]{mirror.pdf} 
%	\caption{The $r_1$, $r_2$ is two separate microphones, and $s_j$ is a sound position on the sound trajectory. The $r_{1,2}$ and $r_{1,3}$ is mirrored microphone positions of $r_1$. The $r_{2,2}$ and $r_{2,3}$ is mirrored microphone positions of $r_2$.}
%	\label{fig::fourier}
%	%\vspace{-3mm}
%\end{figure}

%\begin{figure}[h]
%\centering
%\def\svgwidth{5cm}
%\includegraphics[height=6cm,width= 0.95 \columnwidth]{system_design.pdf}
%\caption{There is time difference of arrival for the same pulse in different microphones due to distance differences between sound and microphones. Here the impulse response describes the reaction of the system as a function of time.}
%\label{fig::system}
%\vspace{-3mm}
%\end{figure}

In the sequel we will use the {\em matching vector} for time
matchings of the same signal at some time instant in each channels,
which is denoted as $(u_1, u_2, \ldots u_m)^T$. We will allow missing data
for such vectors, i.e.\ there might be one or several indices in a vector that has unknown values. 
The components of the vector might also contain outliers. 
Also introduce $o = c (t_1-t_0) = || \bf  \rrr_1 - s ||$ as the offset. This can be interpreted as the distance from the sound to microphone $1$. 
Using this notation the measurement equation (1) becomes 
$$ u_i = || {\bf  \rrr_i - \sss}|| - o.$$
Let $j$ be used as an index for different sounds.
The key idea is that using a number of such measurements $u_{ij}$ it is possible to estimate the
unknown parameters $({\bf \rrr_i, \sss_j}, o_j)$ so that
$$ u_{i,j} = || {\bf \rrr_i - \sss_j }|| - o_j.$$

\section{Data collection}
\label{sec:data}

We have made several experiments with 8 microphones. The measured data (i.e.\ received signals at the microphone array) was obtained by 8 microphones (Shure SV100 or T-bone MM-1) which are connected to an audio interface (M-Audio Fast Track Ultra 8R), connected to a laptop. The microphones were positioned in a room with approximate distance \mbox{0 - 3} meters from each other. 
%For the system we assume that
%the microphones can be in unknown general 3D positions and that the microphones are synchronized. We do not
%put any constraints on the sound sources. They can also be
%in general 3D positions and do not have to fulfill the far-field
%criteria. There can be multiple moving sound
%sources. 
The input to the system is a sound recording with $M$ channels (one from each microphone).%, see Figure~\ref{fig::system}. 
% any??

The $M$ sound channels were sampled at $96000$ Hz. We assume speed of sound is approximately $c = 343 m/s$ for room temperature and normal atmospheric pressure.

Sounds can be generated in several scenarios, for e.g.:
\begin{itemize}
	\item Random distinct sound bursts made for example by banging two spoons together. This produces a set of discrete sound events that are relatively easy to detect and match. 
	\item One continuously moving sound source playing part of a song. This produces a set of smoothly changing time-differences. 
	\item Several continuously moving sound sources.
	\item Mixture of several people talking, clapping, walking around in the room. The sound sources appear and disappear, humans start and stop talking.
	\item In anaechoic chamber
	\item Lunch room.
	\item Lecture room.
\end{itemize}

%Below we give introduction to our data set which we mostly worked on in this thesis work, more information about data set can be seen on Table~\ref{table:data}. 

%\begin{itemize}
%\item Data set \Rmnum{1}: Part of a choir song played by a mobile phone through a small speaker. The sound source is moving slowly through the room. 
%\item Data set \Rmnum{2}: Part of a punk song played by a mobile phone through a small speaker. The sound source is moving slowly through the room. The sound source path also goes through the microphone cluster. 
%\item Data set \Rmnum{3}: Part of a ? song played by a mobile phone through a small speaker. The sound source is moving slowly through the room and at the same time another person make the video. 
%\item Data set \Rmnum{4}: Part of a ? song played by a mobile phone through a small speaker. The sound source is moving slowly through the room, and others make the videos. 
%\end{itemize}

\begin{figure}[h]
    \centering
    \begin{subfigure}[b]{0.45\columnwidth}
        \includegraphics[width=\textwidth]{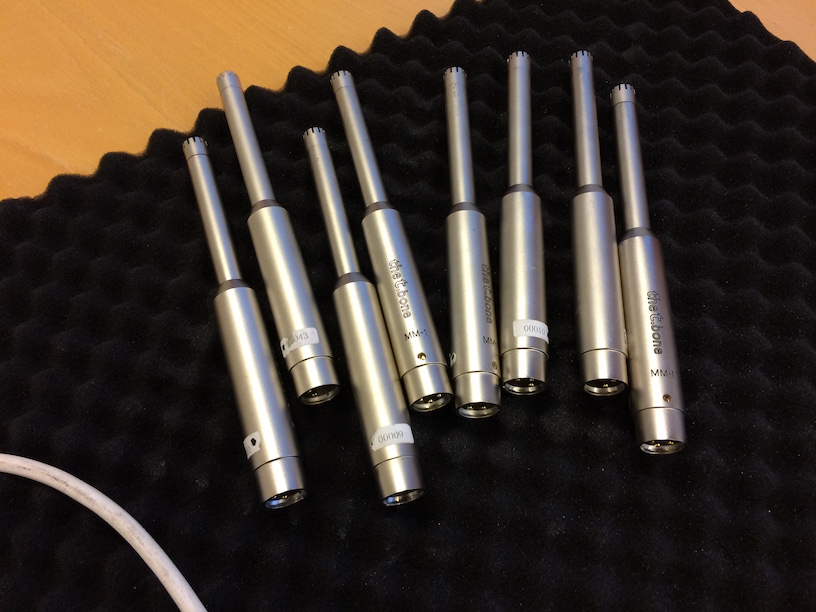}
        \caption{Microphones.}
        \label{fig:gull}
    \end{subfigure}
    ~ %add desired spacing between images, e. g. ~, \quad, \qquad, \hfill etc. 
    \begin{subfigure}[b]{0.45\columnwidth}
        \includegraphics[width=\textwidth]{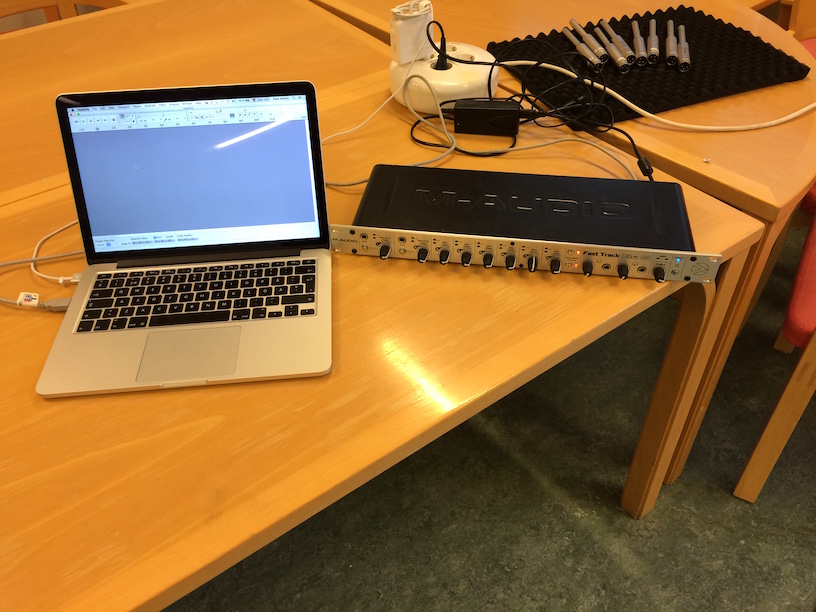}
        \caption{Computer and audio interface.}
        \label{fig:tiger}
        \end{subfigure}
       \caption{Hardwares for data collection.}	\label{fig::filter1}
       \end{figure} 
       
For the datasets, we have also obtained ground truth microphone positions and sound source motion. 

%In the following chapters, we illustrate some of the steps of the automatic system with several experiments. 

%For clarity, here we provide  Table~\ref{table:data}. 

%\begin{table}
%\begin{tabular}
%{ |p{2.4cm}||p{1cm}|p{1cm}|p{1cm}|p{1cm} | }
%%\hline
%%\multicolumn{3}{|c|}{Country List} \\
%\hline
%%\romannumeral 2
% Data set & \Rmnum{1} & \Rmnum{2}& \Rmnum{3} & \Rmnum{4}\\
%\hline
%Nr. microphones & 8   & 8 & 8 &8 \\
%\hline
%Nr. sound positions & 00 & 00 &  00 & 00 \\
%\hline
%Room environment &echo-free-room  &echo-free-room  &Dining area & Lecture room\\
%\hline
%Video clip &No  & No &Yes  & Yes\\
%\hline
%\end{tabular}
%\caption{Information about data set. \label{table:data}}
% \end{table}
 
\section{Time-Difference Estimation and Tracking}
\label{sec:tracking}

Experiments were made in different environments (normal and echo-free). Different types of sound sources were used (claps, voices, continuous songs). Continuously moving sound sources are substantially more challenging. 
%Here we have studied the case of a single moving sound source, but also multiple moving sound sources.

%The current system uses two types of signal processing components:\\
%\indent 1. For claps, we use flank detectors that detect the onset times $t_{ij}$ of such claps. We then use a simple matching scheme to match the onset times along the different sound channels. Here we assume that the different claps do not occur too frequently. \\
%\indent 2. For other sound sources, we search for time differences between channel 1 and channel $i$ using different errors measures. We have used GCC-PHAT (Generalized Cross Correlation with Phase Transform), normalized cross-correlation and similar techniques.\\

For both signal processing components the output is an estimate of range difference $w_{i_1,k_1,i_2,k_2}$ for the direct path $(k_1,k_2)=(1,1)$ at a number of time instants $t_j, j = 1, \ldots n$.

\subsection{Time-Difference Estimation - Claps}

For claps, we use flank detectors that detect the onset times $t_{i}$ of such claps for each channel $i$. We then use a simple matching scheme to match the onset times along the different sound channels. Here we assume that the different claps do not occur too frequently.

Assume that there are $n$ claps, and that each clap is index by $1\le j \le n$. For each such clap we thus get $m$ onset times $t_1, \ldots, t_m$ for the $m$ channels. Assuming that the onset times correspond to the arrival of the direct component of the sound wave we thus get range difference estimates 
\begin{equation}
w_{1,1,i,1} = v (t_i-t_1) .
\label{eq:clap}
\end{equation}

Introduce the column vector $\mathbf{u} \in \fatR^m$, whose components are
\begin{equation}
u_{i} = w_{1,1,i,1} = v (t_i-t_1) . 
\end{equation}
Notice that by definition $u_1 = 0$. 
Each sound clap $j$ gives rise to such a different matching vector $\mathbf{u}_j$.
Now define the $m \times n$ matrix
\begin{equation}
\mathbf{U} =  \begin{pmatrix} \mathbf{u}_1 & \cdots & \mathbf{u}_n 
\end{pmatrix} ,
\end{equation}
whose columns correspond to the matching vectors of the $n$ claps. 
Each sound clap $j$ has a 3D position vector $\mathbf{s}_j$. 
Notice that 
\begin{equation}
\mathbf{U}_{ij} =  || \rrr_{i,1}  - \sss_j || + o_j ,
\end{equation}
with $o_j = - || \rrr_{1,1} - \sss_j  ||$.
The estimation of $o_j$, $\sss$ and $\rrr$ are discussed in Sections~\ref{sec:o} and \ref{sec:rs}.

\subsection{Time-Difference Estimation - Continuously moving sound sources}

\begin{figure}
  %\centering
  \setlength\figureheight{0.4\linewidth}
  \setlength\figurewidth{0.8\linewidth}
  \tikzsetnextfilename{gcc}
% Minimal pgfplots version: 1.3
%  http://www.mathworks.com/matlabcentral/fileexchange/22022-matlab2tikz
\begin{tikzpicture}

\begin{axis}[%
width=0.95092\figurewidth,
height=\figureheight,
at={(0\figurewidth,0\figureheight)},
scale only axis,
axis on top,
every outer x axis line/.append style={black},
every x tick label/.append style={font=\color{black}},
xmin=1,
xmax=3466,
xtick={1,481,961,1441,1921,2401,2881,3361},
xticklabels={{0},{5},{10},{15},{20},{25},{30},{35}},
xlabel={\xlabelfortikz},
every outer y axis line/.append style={black},
every y tick label/.append style={font=\color{black}},
y dir=reverse,
ymin=1,
ymax=1601,
ytick={236,518,801,1083,1365},
yticklabels={{-2},{-1},{0},{1},{2}},
ylabel={\ylabelfortikz},
title={\titlefortikz},
axis x line*=bottom,
axis y line*=left,
x unit=s,change x base=true,y unit=m,change y base=true
]
\addplot [forget plot] graphics [xmin=0.5,xmax=3466.5,ymin=0.5,ymax=1601.5] {figures/gcc-1-mod-lowres.jpg};
\end{axis}
\end{tikzpicture}%
%  \vspace*{-4mm}
  \caption{GCC-PHAT between all frames for two channels (channel 1 vs channel 2). Each pixel in column represents one correlation value of frames from channel 1 and 2. Only positive values plotted. The most prominent curve structure is the result fo the direct sound path. }\label{fig:gcc}
\end{figure}
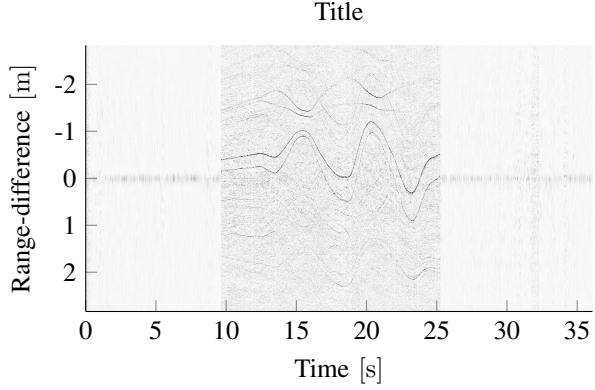

Each sound file is divided into frames, 2048 sample points long, 1000 sample points apart. %
The Generalized Cross Correlation with Phase Transform (GCC-PHAT) is calculated between corresponding frames for all microphone pairs. %
Arranging the correlation score of each frame as a column in a matrix yields a score matrix as illustrated in Figure~\ref{fig:gcc}. In the figure the x-axis represents time into the recording in seconds and the y-axis are range-differences $v(\tau_{m_2}-\tau_{m_1})$ in meters. As can be seen in the figure the data is quite noisy. Nevertheless, certain patterns in the data can be discerned. These curves correspond to direct and indirect range-differences 
$w_{i_1,k_1,i_2,k_2} = || \sss - \rrr_{i_2,k_2} || - || \sss - \rrr_{i_1,k_1} ||$, for different combinations of $k_1$ and $k_2$, 
see Equation~\ref{eq:uikik}.

In general, the received signal model of two microphones can be expressed as
$$f_1(t) = g(t) + n_1(t),$$
$$f_2(t) = g(t-\tau)+n_2(t),$$
where $n_1(t)$ and $n_2(t)$ are noise and $\tau$ is the time difference of the sound source to the microphone pair. 
The generalized cross-correlation or GCC function between $f_1(t)$ and $f_2(t)$ is 
$$ R_{y_1y_2}(\tau) = \int_{-\infty}^{\infty}W(\omega)G_{f_1f_2}(\omega)e^{j2\pi \omega \tau}d\omega,$$
where $G_{f_1f_2}$ is the cross-power spectrum of received signals of two microphones, and the generalized frequency weighting is
$$W(\omega) = H_1(\omega)H_2^*(\omega) = \frac{1}{|G_{f_1f_2}(\omega)|},$$
where * denotes the complex conjugate. 
%In practice, only estimate of the cross-power spectral density can be obtained from finite observations of the received signals. Consequently, the integral
%$$ \hat{R}_{y_1y_2} (\tau) = \int_{-\infty}^{\infty}W(f)\hat{G}_{f_1f_2}(f)e^{j2\pi \omega\tau}d\omega$$
%is evaluated and used for estimating delay. 
%% Fattar inte den andra integralen. Vad var det för skillnad????

The main idea here is that some of the peaks in Figure~\ref{fig:gcc} correspond to 
range-difference measurements
\begin{equation}
w_{i_1,k_1,i_2,k_2} = || \sss - \rrr_{i_2,k_2} || - || \sss - \rrr_{i_1,k_1} || .
\end{equation}
The difficulty lies in determining which peaks are relevant 
and for each such peak to determine the corresponding multi-path index pair $(k_1,k_2)$. 
If such correspondences can be found the next step is the geometric estimation of microphone and sound source positions.

The strategy used here is to use heuristics to try to first find peaks that correspond to the direct path components, i.e.\ those with multi-path index pair $(k_1,k_2) = (1,1)$. 

\subsection{Peak selection}

The four strongest peaks for each frame from the cross-correlation are selected. %, see Figure~\ref{fig:toppeaks}. %
Only peaks above a certain threshold are considered to reduce the number of outliers. Many of these peaks correspond to direct path range differences $w_{i_1,1,i_2,1}$. For a continuously moving sound source these peaks typically generate smooth tracks in the figure. 
As can be seen in the figure, the set of detections contain a mixture of true positives and false positives. The false positives consists of both noise but also of range difference for multi-path components, i.e.\ 
range differences $w_{i_1,k_1,i_2,k_2}$ with multi-path index pair $(k_1,k_2) \neq (1,1)$.
The number of false positives are evaluated in the experimental evaluation, (Section \ref{sec:falsepositives}).

\begin{figure}
  %\centering
  \setlength\figureheight{0.3\linewidth}
  \setlength\figurewidth{0.5\linewidth}
  \tikzsetnextfilename{matchedpeaks}
  \input{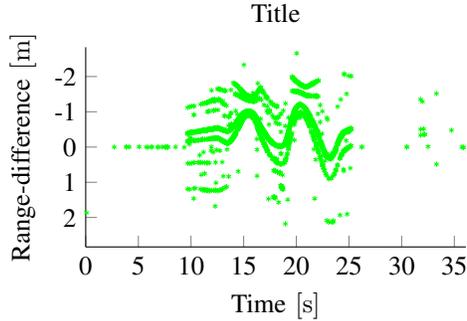}
%  \vspace*{-4mm}
  \caption{Peaks after matching using data from all channels.}\label{fig:matchedpeaks}
\end{figure}

\subsection{Peak tracking}

\begin{figure}
  %\centering
  \setlength\figureheight{0.3\linewidth}
  \setlength\figurewidth{0.6\linewidth}
  \renewcommand{\titlefortikz}{Peaks connected with RANSAC}
  \tikzsetnextfilename{ransac}
  \input{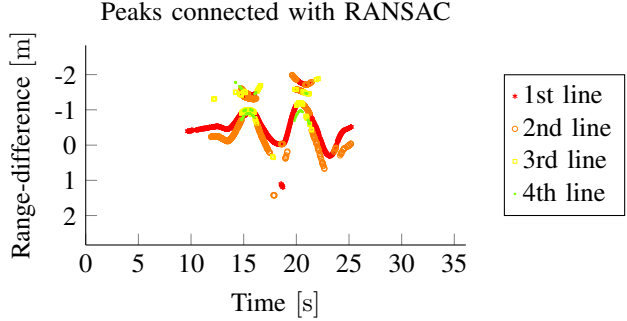}
%  \vspace*{-4mm}
  \caption{Inliers for line segments found with RANSAC.}\label{fig:ransac}
\end{figure}

Assuming that the motion of the sound source is relatively smooth (at least at times), we exploit the continuity constraint by grouping together range difference peaks. Here we use a non-repeating 
RANSAC algorithm to find so called {\em tracklets}, i.e.\ small track parts of the curves.
These are 21 frames wide with 1 frame overlap.
Peaks closer than a threshold from the line are considered to be inliers.
Both the lines with the most inliers and the inliers themselves are stored for each tracklet. 
If two lines share more than one inlier, the one with less inliers is ignored. The result is illustrated in Figure~\ref{fig:ransac}. Notice that the process further removes outlier and errors while keeping the dominant
range difference tracks in the data. 

In the next step, all line segments closer than a certain threshold are connected to form longer tracks. 
% %segment är nog fel ord
%, see Figure~\ref{fig:segments}. %
%
%\begin{figure}
%  %\centering
%  \setlength\figureheight{0.5\linewidth}
%  \setlength\figurewidth{0.9\linewidth}
%  \renewcommand{\titlefortikz}{Connected delay segments}
%  \tikzsetnextfilename{segments}
%  \input{figures/segments.tikz}
%  \vspace*{-6mm}
%  \caption{Connected line segments.}\label{fig:segments}
%\end{figure}
%
Segments separated by a small amount in time sharing roughly the same line equation are connected to form even longer segments. 
The hypothesis is that the longest segment corresponds to the direct path. %, see Figure~\ref{fig:connectedsegments}.

Thus we obtain an hypothesis of the range difference for the direct path $w_{i_1,1,i_2,1}$ at a number of instances $t_j, j = 1, \ldots, n$ along the time axis.  
%Similar to the case of claps we collect the range differences to microphone $1$ in matching vectors for each time instance $t_j$. 
For each time instance $t_j$ we collect the range difference estimates $w_{1,1,i,1}$ in a 
column vector, $\mathbf{u} \in \fatR^m$, i.e.\ so that the components are given by
\begin{equation}
u_{i} = w_{1,1,i,1} .
\end{equation}
Now define the $m \times n$ matrix
\begin{equation}
\mathbf{U} =  \begin{pmatrix} \mathbf{u}_1 & \cdots & \mathbf{u}_n 
\end{pmatrix} ,
\end{equation}
whose columns correspond to the matching vectors of the $n$ time instants $t_j$. 
Again notice that 
\begin{equation}
\mathbf{U}_{ij} =  || \sss_j - \rrr_{i,1} || + o_j ,
\end{equation}
with $o_j = - || \sss_j - \rrr_{1,1} ||$.
The estimation of $o_j$, $\sss$ and $\rrr$ are discussed in Sections~\ref{sec:o} and \ref{sec:rs}.
%where $\sss_j$ denotes the 3D position of the sound source at time $t_j$.
%The estimation of $\sss$ and $\rrr$ are discussed in Sections~\ref{sec:xxx}.

\begin{figure}
  %\centering
  \setlength\figureheight{0.3\linewidth}
  \setlength\figurewidth{0.6\linewidth}
  \renewcommand{\titlefortikz}{Smoothed delays}
  \tikzsetnextfilename{smooth}
  \input{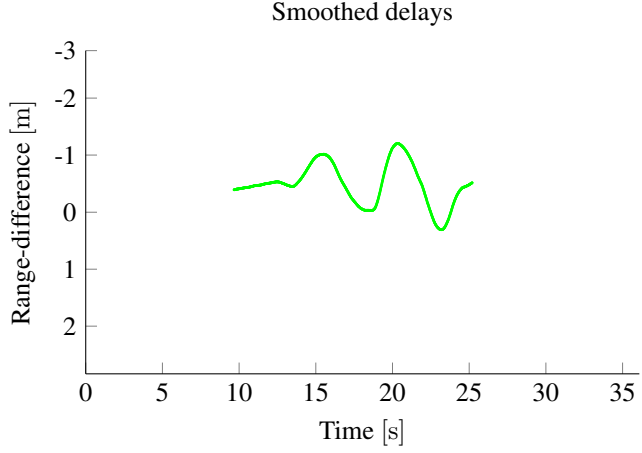}
  \caption{Peaks after smoothing.}\label{fig:smooth}
\end{figure}

\section{Offset Estimation}
\label{sec:o}

From the signal processing step we obtain a $m \times n$ matrix $\mathbf{U}$
of range difference estimates $\mathbf{U}_{ij}$. These are linked to the 
positions of the $m$ microphones $\rrr_{1,1}, \ldots, \rrr_{m,1}$, 
the $n$ sound events $\sss_1, \ldots, \sss_n$ and the $n$ offsets $o_1, \ldots, o_n$ according to 
\begin{equation}
\mathbf{U}_{ij} =  || \sss_j - \rrr_{i,1} || + o_j , 
\end{equation}
where the offsets are defined as
\begin{equation}
o_j=   - || \sss_j - \rrr_1 || .
\end{equation}
In order not to clutter the notation in this section we are going to drop the multi-path index, i.e.\
write $\rrr_i= \rrr_{i,1}$. 

In this section we derive the rank constraint, which can be used for 
estimating the offsets $o_j$ from the elements of matching vectors $\mathbf{U}_{ij}$. 
This is an extension of the work in \cite{jiang-etal-icassp13,kuang-astrom-2espc2-13}. 
%The strategy here has been %to develop algorithms
%\begin{itemize}
%\item A: Rewrite equation%Calculating the offsets from minimal data
%\item B: Factorization%Iterative optimization methods for solving
%\item C: Solve for the unknown transformation.
%\end{itemize}

\subsection{The Rank Constraint}

To begin with, let us rewrite the measurement equation 
\begin{equation}
\mathbf{U}_{ij} =  || \sss_j - \rrr_i || + o_j .
\end{equation}
as
$$ (\mathbf{U}_{ij}-o_{j})^2 = ||\bf \rrr_i -  \sss_j ||^2 = \rrr_i^{T}\rrr_i - 2\rrr_i^{T}\sss_j + \sss_j^{T}\sss_j . $$
A simple rearrangement yields
$$ \mathbf{U}_{ij}^2 - 2\mathbf{U}_{ij}o_{j}^2 =  \rrr_i^{T}\rrr_i - 2\rrr_i^{T}\sss_j + ( \sss_j^{T} \sss_j - o_j^2). $$
By constructing the vectors
%\begin{center}
$ \tilde{\mathbf{R}}_i =  \left[ \begin{matrix}  1  &  \mathbf{r}_i^T   &\mathbf{r}_i^T \mathbf{r}_i  \end{matrix} \right]^T $ and 
$ \tilde{\mathbf{S}}_j=   \left[ \begin{matrix}  \mathbf{s}_j^T \mathbf{s}_j - o_j^2 &  -2 \mathbf{s}_j^T   & 1    \end{matrix} \right]^T  $,
%\end{center}
we obtain 
\begin{equation}
 \mathbf{U}_{ij}^2 - 2\mathbf{U}_{ij}o_{j}^2  =  \tilde{\mathbf{R}}_i^T \tilde{\mathbf{S}}_j .
\end{equation}
By collecting $\tilde{\mathbf{R}}_i$ and $\tilde{\mathbf{S}}_j$ into the matrices $\tilde{\mathbf{R}} \in \mathbb{R}^{5 \times m}$ and $\tilde{\mathbf{S}} \in \mathbb{R}^{5 \times n}$, we have $ \mathbf{D} =\tilde{ \mathbf{R}}^T\tilde{\mathbf{S}}$, where $\mathbf{D}$ is the $m \times n$  matrix with elements  $d_{ij}=  u_{ij}^2 - 2 u_{ij}o_j $. This gives a matrix $\mathbf{D}$ that is at most of rank $5$ as we increase $m$ and $n$. 

Now form the matrix $ \mathbf{F}= \frac{1}{-2} \mathbf{C}_M^{T} \mathbf{D} \mathbf{C}_N$,
where
$\mathbf{C}_m = \left[ -\mathds{1}_{m-1} \ \mathbf{I}_{m-1} \right]^T $,
$\mathbf{C}_n = \left[ -\mathds{1}_{n-1}  \ \mathbf{I}_{n-1} \right]^T$,
and
where $\mathds{1}_{n-1}$ is a $ (n-1) \times 1$ vector with $1$ as entries and $\mathbf{I}_{n-1}$ is identity matrix of size ($n-1$). 
This matrix $\mathbf{F}$ has size $(m-1) \times (n-1)$. The elements are $ f_{i,j} = \frac{1}{-2} (\tilde{d}_{i,j} - \tilde{d}_{i,1} - \tilde{d}_{1,j} + d_{1,1} )$ as shown below.

%\vspace{5mm}
%\resizebox{0.95\linewidth}{!}{
%%\begin{tiny}
%%\begin{align}
%$
%%\[ 
%F =
%\begin{pmatrix}
% -1 & -1 & \cdots & -1  \\
%  1 &    &  &    \\
% &   1 &    &    \\
% &    &     &\ddots \\
% &    &     & 1 
%\end{pmatrix}
%\begin{pmatrix}
%d_{11} & d_{12} & \cdots & d_{1N} \\
%  d_{21} & d_{22} & \cdots & d_{2N} \\
%  \vdots  & \vdots  & \ddots & \vdots  \\
%d_{M1} & d_{M2} & \cdots &d_{MN}
%\end{pmatrix}
%\begin{pmatrix}
%       -1 & 1 &    &         \\
%       -1 &    & 1 &       \\
%\vdots &    &    & \ddots   \\
%       -1 &    &     &    & 1
%\end{pmatrix}\nonumber
%$
%}
%\\
%\resizebox{0.95\linewidth}{!}{
%$\hspace{1.0 cm}=
%\begin{pmatrix}
%-d_{11} - d_{21} \cdots -d_{M1}&-d_{12} - d_{22} \cdots -d_{M2}& \cdots & -d_{1N} - d_{2N} \cdots -d_{MN} \\
%d_{11} & d_{12}  & \cdots & d_{1N} \\
%  \vdots  & \vdots  & \ddots & \vdots  \\
%d_{M-1,1} & d_{M-1,2}  & \cdots & d_{M-1,N}
%\end{pmatrix}
%\begin{pmatrix}
%       -1 & 1 &    &         \\
%       -1 &    & 1 &       \\
%\vdots &    &    & \ddots   \\
%       -1 &    &     &    & 1
%\end{pmatrix}\nonumber
%%\]
%$
%%\end{align}
%%\end{tiny}
%}

\vspace{5mm}
\resizebox{0.95\linewidth}{!}{
%\begin{tiny}
%\begin{align}
$
%\[ 
\mathbf{F} =\frac{1}{-2} 
\begin{pmatrix}
       -1 & 1 &    &         \\
       -1 &    & 1 &       \\
\vdots &    &    & \ddots   \\
       -1 &    &     &    & 1
\end{pmatrix}
\begin{pmatrix}
d_{11} & d_{12} & \cdots & d_{1n} \\
  d_{21} & d_{22} & \cdots & d_{2n} \\
  \vdots  & \vdots  & \ddots & \vdots  \\
d_{m1} & d_{m2} & \cdots &d_{mn}
\end{pmatrix}
\begin{pmatrix}
 -1 & -1 & \cdots & -1  \\
  1 &    &  &    \\
 &   1 &    &    \\
 &    &     &\ddots \\
 &    &     & 1 
\end{pmatrix}\nonumber
$
}\vskip 3mm
\resizebox{0.95\linewidth}{!}{
$\hspace{1.0 cm}= \frac{1}{-2} 
\begin{pmatrix}
-d_{11} + d_{21} & -d_{12} + d_{22} & \cdots & -d_{1n} + d_{2n} \\
%-d_{11} + d_{31} & -d_{12} + d_{32} & \cdots & -d_{1n} + d_{3n}  \\
  \vdots  & \vdots  & \ddots & \vdots  \\
-d_{11} + d_{m1} & -d_{12} + d_{m2} & \cdots & -d_{1n} + d_{mn} 
\end{pmatrix}
\begin{pmatrix}
 -1 & -1 & \cdots & -1  \\
  1 &    &  &    \\
 &   1 &    &    \\
 &    &     &\ddots \\
 &    &     & 1 
\end{pmatrix}\nonumber
%\]
$
%\end{align}
%\end{tiny}
}\vskip 3mm
\resizebox{0.95\linewidth}{!}{
$\hspace{1.2 cm}= \frac{1}{-2} 
\begin{pmatrix}
d_{11} - d_{21} - d_{12} + d_{22}&  \cdots & d_{11} - d_{21} - d_{1n} + d_{2n}\\
%d_{11} - d_{31} - d_{12} + d_{32}&  \cdots & d_{11} - d_{31} - d_{1n} + d_{3n}  \\
  \vdots  & \ddots & \vdots  \\
d_{11} - d_{m1} - d_{12} + d_{n2}&  \cdots & d_{11} - d_{m1} - d_{1n} + d_{mn}
\end{pmatrix}
%\begin{pmatrix}
%d_{11} - d_{21} - d_{12} + d_{22}& d_{11} - d_{21} - d_{13} + d_{23}& \cdots & d_{11} - d_{21} - d_{1N} + d_{2N}\\
%%d_{11} - d_{31} - d_{12} + d_{32}& d_{11} - d_{31} - d_{13} + d_{33}& \cdots & d_{11} - d_{31} - d_{1N} + d_{3N}  \\
%  \vdots  & \vdots  & \ddots & \vdots  \\
%d_{11} - d_{M1} - d_{12} + d_{M2}& d_{11} - d_{M1} - d_{13} + d_{M3}& \cdots & d_{11} - d_{M1} - d_{1N} + d_{MN}
%\end{pmatrix}
$
}
\vspace{5mm}

It is relatively easy to see that $\rank (\mathbf{F}(o)) \leq 3$. In fact
$$ \mathbf{F}(o)=\mathbf{R}^T \mathbf{S} , $$
with 
$$\mathbf{R} = \begin{bmatrix} \mathbf{r}_2-\mathbf{r}_1, \hdots, \mathbf{r}_m-\mathbf{r}_1 \end{bmatrix}$$ and 
$$\mathbf{S} = \begin{bmatrix} \mathbf{s}_2-\mathbf{s}_1, \hdots, \mathbf{s}_n-\mathbf{s}_1 \end{bmatrix}.$$
The matrix $\mathbf{F}(o)$, here called the {\em double compaction matrix}, depends on the matching vectors $\mathbf{U}_{j}$ and on the offsets $o_j$. 
Notice here that if the dimension of the affine span of either the microphones or the the sound events are less than 3, then the rank will be lowered. For example if all the receivers lie in a plane, then $\rank \mathbf{R} \leq 2$ and consequently $\rank \mathbf{F} \leq 2$. Another example is if all the sound events are on a line then $\rank \mathbf{S} \leq 1$ and consequently $\rank \mathbf{F} \leq 1$. 

%\begin{figure}
%\centering
%\def\svgwidth{5cm}
%\includegraphics[height=3cm,width=8cm,trim = 30mm 60mm 20mm 60mm]{ransac2.pdf}
%\caption{The robust parameter estimation is based on the RANSAC paradigm. In the illustration there are 8 channels (rows) and 56 (out of 129) matching vectors (columns), we choose the smaller set of matching vectors only for better display reason. A random subset of 7 channels and 6 matching vectors (illustrated in blue) are used to estimate the parameters. The remaining data is used to verify (or falsify) the estimate. For this starting point there are a substantial number of inliers (illustrated with green) indicating that this is indeed a promising estimate. Outliers are shown in red and missing data are shown in black.}
%\label{fig::ransac}
%\end{figure}

\subsection{Minimal solvers for the rank constraint}

The problem of determining the offsets $o$ now becomes that of finding $o$ so that $\rank (\mathbf{F}(o)) \leq 3$.
The elements of $\mathbf{F}$ are linear in $o$. The fact that the rank is less than $3$ means that all $4 \times 4$ sub-determinants of $\mathbf{F}$  are zero. These constraints on these sub-determinants or minors are thus 
polynomial constraints of degree $4$ in $o$. 

For a $(m-1) \times (n-1)$ matrix 
$\tilde{\mathbf{F}} \ $of rank $3$,  the number of constraints $N_c = |\Lambda_{4}| = \left( \begin{matrix} m-1 \\ 4 \end{matrix} \right) \left( \begin{matrix} n-1 \\ 4 \end{matrix} \right)$ among which $(m-4)(n-4)$ constraints are linearly independent. Each constraint is a polynomial equation of degree $4$ in $\{o_1,\dots,o_n\}$. For different choices of $m$ and $n$,  this system of polynomials equations can either be well-defined, over-determined or under-determined. To resolve this, we rely on algebraic geometry tools and make use of \textsl{Macaulay2} \cite{M2}.

%Here it is interesting to note that the problem becomes fundamentally different if either the microphones or the sound sources lie in a plane. In this case the rank of matrix $\tilde{\mathbf{F}}$ is at most 2. Similarily, if either the microphones or the sound sources lie in a line then the rank of matrix $\tilde{\mathbf{F}}$ is at most 1.

It turns out that there are several choices for $m$ and $n$ that produce well-defined and solvable polynomial systems. We summarize those cases and the number of solutions of the related polynomial systems for $K = 3$ and $ K = 2$ in Table \ref{table_case}. In the following discussion, we denote the case with $m$ receivers and $n$ transmitters as $m$r/$n$s. 
Given these solvable cases, we can apply numerically stable polynomial solvers based on methods described in \cite {byrod-josephson-etal-ijcv-09} to solve for the unknown offsets. Using this technique, solving the system of polynomial equations is converted to an eigenvalue problem, which is efficiently solved by techniques from numerical linear algebra. The resulting solvers are fast (execution time in the order of milliseconds) and are non-iterative (in the sense that they only use standard numerical linear algebra routines and no iterations). 

\begin{table}
\centering
\begin{tabular}{c c c c }
\hline\hline
$K$ & $m$ & $n$ & $N_{sol}$ \\ [0.5ex] % inserts table %heading
\hline
3 & 9 & 5 & 1 \\
3 & 7 & 6 & 5 \\
\vspace{-.1 mm}3 & 6 & 8 & 14  \\ [1.3mm]
\hline  \\[-1.8mm]
2 & 7& 4 & 1   \\
2 & 5 & 6 & 5  \\ 
\hline
\end{tabular}
\caption{Number of solutions to the polynomial systems of rank constraints on unknown offsets for different cases in (III) (3D and 2D). 
%For linear cases in (II), the number solution is $1$
}
 \label{table_case}
%\end{table}
\end{table}

%For most of the experiments we have assumed the most general case, i.e.\ that both microphones and sound sources span 3D. In this case there are three minimal problems for determining offsets so that the double compaction matrix $\mathbf{F}(\mathbf{U},\mathbf{o})$ 
%has rank 3. The minimal problems are
%\begin{itemize}
%\item 9 microphones and 5 sounds (unique solutions).
%\item 7 microphones and 6 sounds (five solutions).
%\item 6 microphones and 8 sounds (14 solutions).
%\end{itemize}
%For all of these problems there are efficient closed form algorithms for finding all solutions, cf.\ \cite{kuang-astrom-2espc2-13}. 

\subsection{Robust estimation of offsets}

The resulting matching $\mathbf{U}$ from the signal processing steps inlude (i) inlier measurements with noise, (ii) outliers due to gross errors in matching and (iii) missing data. 
Robust estimation of offsets can be achieved by algorithms that try to find the offsets with the most inliers (or the fewest outliers), for this propose we use RANSAC, cf.\ \cite{fischler-bolles-ca-81}. It has proved to be useful for many difficult parameter estimation problems in signal processing and computer vision, e.g.\ \cite{schmalenstroeer2011unsupervised,valente2010self,batstone_icc_2016}, since the methods are robust to both measurement errors and outliers. RANSAC select the solution with the most inliers and among these the one with the lowest error score. The algorithm requires two parameters, the agreement threshold (how close does an inlier have to be) and number of iterations.

The main idea of RANSAC is to use an hypothesize and test paradigm. First one hypothesize a minimal subset of data for which it is possible to solve for the parameters (the offsets in this case). Using this hypothesized solution one then extends the solution to the rest of the data and determines how many of the remaining data that agrees with the solution. Here it is important that both the minimal solvers and the verification step is fast, since this allows for more iterations. 
All of the minimal solvers that we have generated (Table~\ref{table_case}) are potentially useful. Here we explain the methodology using the 7 microphone, 6 sound events (rank 3) solver. 

\begin{enumerate}
\item From the set of matching vectors, randomly select a subset of 7 channels and 6 matching vectors, and solve for the offset for the minimal case $7 \times 6$ above. Use the closed form algorithm for finding the offsets ${o_j}$ for these 6 matching vectors. The relevant (valid) solutions should have
offsets that are real, and since $|| {\bf \rrr_i - \sss_j }|| = u_{ij} - o_{j}$, the offsets should fulfill the constraints $u_{ij} \geq o_j$ for $i = 1, \ldots, M$ and $j = 1, \ldots, N$. Ignore solutions that do not fulfill these constraints. 
\item Extend to the remaining microphones, and solve for $M \times 6$.
\item For each solution study how many of the remaining matching vectors that fulfill the geometric constraint, keep largest set of inliers. The measurements treat as inliers if $|\mathbf{d}_{ij} - ||\mathbf{m_i} - \mathbf{s_j}||_2| \le \epsilon$ for a given $\epsilon$. 
\item Repeat (1) to (3) a fixed number of times and choose the solution with the maximum number of inlier matching vectors.
\end{enumerate}

\subsection{Non-linear optimization}

Given an initial estimate of offsets it is possible to improve on the offsets using non-linear optimization. 
%Non-linear optimization has many different implementations and it remains the dominant structure refinement technique for real applications. It is slower but more accurate. 
%Non linear iterative methods are more accurate than linear method, but require first guess, bundle adjustment can use factorization result as the first guess. 
Typically this is applied once a rough initial estimation has been found, such initial estimates may be obtained in a number of different ways including use of minimal solvers in a RANSAC framework, estimation of parameters using factorization techniques, etc. 

\subsection{Rank-based Nonlinear Optimization}

In this section, we derive a iterative nonlinear optimization method for improving the estimate of the unknown offsets $o_j$. 
While the non-iterative schemes presented above apply only to specific number of $M$ and $N$ with no missing data, the method present in this section copes with such cases naturally. 

Given the knowledge that the measurement matrix after compaction is of rank $K$, we can derive another scheme based nonlinear optimization to estimate the offsets  $\mathbf{o} = (o_1,o_2,\ldots, o_N)$. The idea is to find the offset such that the measurement matrix after compaction is as close to a rank-$K$ matrix as possible. Thus, we have the following minimization problem: 
\begin{eqnarray} 
\min_{\mathbf{o}, \mathbf{A}} {||\mathbf{F}(\mathbf{U},\mathbf{o}) - \mathbf{A}||}_{F,\Omega} , \nonumber \\
s.t. \ \rank (\mathbf{A}) = K ,
\label{eq:optrank1} 
\end{eqnarray}
where $\mathbf{F}$ is the matrix resulting from the compaction operators as in the previous section, $\mathbf{U}$ is matching matrix, $\mathbf{A} \in \mathbb{R}^{(m-1)\times(n-1)}$, and ${||\cdot||}_{F,\Omega}$ is the Frobenious norm on the matrix entries that are observed specified by the set $\Omega$. 

An alternative formulation is to optimize over the unknown offsets and model the errors direction on the measurement matrix as follows.
\begin{eqnarray}
&\min_{\mathbf{o}, \tilde{\mathbf{U}}} & {||\mathbf{U}-\tilde{\mathbf{U}}||}_{F,\Omega} , \nonumber \\
&s.t. &\ \rank \mathbf{F}(\tilde{\mathbf{U}},\mathbf{o}) = K . \nonumber \\
\end{eqnarray}

Here $\tilde{\mathbf{U}}$ is approximation of matching matrix.
Similar formulation of the minimization problem (\ref{eq:optrank1}) has been proposed in \cite{Gaubitch-etal-icassp13} to utilize the rank constraints. Given that rank constraint on $\mathbf{A}$, the minimization problem is non-convex. In \cite{Gaubitch-etal-icassp13}, an alternating scheme is proposed. To be more specific, one first fixes the offsets $\mathbf{o}$, and solve for the optimal $\mathbf{A}$ using SVD. Then, one fixes $\mathbf{A}$, the problem of finding the optimal $\mathbf{o}$ is convex. However, the rate of convergence of this alternative scheme is very slow. Thus, an additional regularization term on $\mathbf{A}$ is introduced to speed up the convergence. Here, we used a gradient descent scheme that utilize a local parameterization of the rank constraints on $\mathbf{A}$ directly. 

%TODO: write about the parameterization

%Simplification of complex processes based on some assumptions. It is difficult to find a minimal and global parameterization of rank $k$ matrices. 
The optimization problem is basically a non-linear least squares problem. Here we adopt a strategy to make local parametrization at each iteration step. For the optimization problem (\ref{eq:optrank1}), this is implemented as follows. Assuming that the estimate for the current iteration $(\mathbf{o}_k, \mathbf{A}_k)$. A local parametrization for the offsets is straightforward. Introduce a local perturbation $(y_1, \ldots, y_n)$, where $x_i$ denotes the change in offset number $i$. For the local parameterization of $\mathbf{A}_k$, we use singular value decomposition to rewrite 
$\mathbf{A}_k$
as
$$\mathbf{A}_k  = U E_K W , $$
where $U$ is the $m \times m$ unitary matrix from the singular value decomposition $\mathbf{A}_k = U S V^T$,
where $W$ is the $K \times n$ matrix corresponding to the first $K$rows of $S V^T$ and
where $E_K$ is a $m \times K$ matrix, the first $K$columns of a $m \times m$ identity matrix. 
A local parametrization is then 
$$\mathbf{A}_{k+1} = U e^{\sum_i^n z_i B_i}E_K (\sum_{j=1}^{Kn} (W + w_j C_j) . $$
Here $C_j$ are a basis for matrices of size $K \times n$, where $B_i$ are a basis for $m \times m$ anti-symmetric matrices with zeros in the upper-left $K \times K$ block and zeros in the lower-right $(m-K) \times (m-K)$ block, as shown below:

\vskip 0.5cm

%\[  \mathbf{B_i} = 
%\left[ \begin{array}{c|c}0_{k \times k}& \times \\ \hline \times &0_{(m-k) \times (m-k)}\end{array}\right]
%\]

%\begin{tikzpicture}[
%style1/.style={
%  matrix of math nodes,
%  every node/.append style={text width=#1,align=center,minimum height=5ex},
%  nodes in empty cells,
%  left delimiter=[,
%  right delimiter=],
%  },
%style2/.style={
%  matrix of math nodes,
%  every node/.append style={text width=#1,align=center,minimum height=5ex},
%  nodes in empty cells,
%  left delimiter=\lbrace,
%  right delimiter=\rbrace,
%  }
%]
%\matrix[style1=0.45cm] (1mat)
%{
%  & & & & & & \\
%  & & & & & &  \\
%  & & & & &  & \\
%  & & & & & & \\
%  & & & & & & \\
%};
%
%\draw[loosely dashed]
%  (1mat-2-1.south west) -- (1mat-2-7.south east);
%\draw[loosely dashed]
% (1mat-1-3.north east) -- (1mat-5-3.south east);
%
%\node[font=\Large] 
%at ([yshift=10pt]1mat-2-2) {$\mathbf{0}_{K \times K}$};
%\node[font=\Large] 
%  at (1mat-2-5.north east) {$\times$};
%\node[font=\Large] 
%  at ([xshift=-5pt, yshift = -1pt]1mat-4-2) {$\times$};
%\node[font=\Large] 
%  at ([xshift=25pt, yshift = -10pt]1mat-4-4.north east) {$\mathbf{0}_{(m-K) \times (m-K)}$};
%
%\draw[decoration={brace,mirror,raise=12pt},decorate]
%  (1mat-3-1.north west) -- 
%  node[left=15pt] {$m-K$} 
%  (1mat-5-1.south west);
%\draw[decoration={brace,mirror,raise=5pt},decorate]
%  (1mat-5-1.south west) -- 
%  node[below=7pt] {$K$} 
%  (1mat-5-3.south east);
%
%\end{tikzpicture}

The parametrization involves $m + K (m-K) + Kn$ parameters
$$ x = \begin{pmatrix} y \\ z \\ w \end{pmatrix}. $$

It is relatively straightforward to calculate the analytic derivatives of $A_{k+1}$ with respect to $x$. And also the derivatives of $F(\mathbf{U},\mathbf{o})$ with respect to $o$. This makes it possible to calculate the residuals and the jacobean of the non-linear least squares problem and implement an efficient Gauss-Newton optimization scheme for solving (\ref{eq:optrank1}).

We then use the minimal solver on random subsets of
data. If the solution is consistent with many of the remaining
data, we extend the solution and then refine it using the iterative
methods. This makes it possible to handle
both missing data and outliers.

\section{Estimation of Microphone and Sound Source Positions}
\label{sec:rs}

Once we have calibrated the measurement matrix with the offsets $\{ o_j \}$, we can obtain distance measurements 
$d_{ij} = \mathbf{U}_{ij} - o_j$. The measurement equation then becomes
\begin{equation}
d_{ij} = \mathbf{U}_{ij} - o_j=  || \sss_j - \rrr_{i,1} || .
\end{equation}
We then proceed to solve the locations of microphones $\{\mathbf{m}_i\}$ and  sounds $\{\mathbf{s}_j\}$ as a TOA problem. Note that one can only reconstruct locations of microphones and sounds up to Euclidean transformation and mirroring. The TOA-based self-calibration problem studied here corresponding to bipartite graph.

Similar to before, by squaring the measurement equations,
\begin{equation}
	d_{ij}^2= ( \mathbf{r}_{i} - \mathbf{s}_{j} )^T ( \mathbf{r}_{i} - \mathbf{s}_{j} )  = \mathbf{r}_{i}^T\mathbf{r}_{i} + \mathbf{s}_{j} ^T\mathbf{s}_{j}  - 2 \mathbf{r}_{i}^T \mathbf{s}_{j}
\end{equation}
we obtain a set of $mn$ polynomial equations in the unknown. We form
$mn$ new equations by linear combinations of the ones above. In particular we choose equations according to:
\begin{eqnarray}
	&d_{11}^2& = (\mathbf{r}_1 - \mathbf{s}_1) ^T (\mathbf{r}_1 - \mathbf{s}_1), \label{con_1} \\
	&d_{1j}^2 - d_{11}^2 &	=  -2 \mathbf{r}_1 ^T (\mathbf{s}_j - \mathbf{s}_1) + \mathbf{s}_{j}^T\mathbf{s}_{j} - \mathbf{s}_{1} ^T\mathbf{s}_{1}  \label{con_2} \\
	&d_{i1}^2 - d_{11}^2 &	=  -2 (\rrr_i - \mathbf{r}_1)^T \mathbf{s}_1 + \mathbf{r}_{i}^T\mathbf{r}_{i} - \mathbf{r}_{1} ^T\mathbf{r}_{1}, \label{con_3} \\
	& \frac{d_{ij}^2 - d_{i1}^2 - d_{1j}^2 + d_{11}^2}{-2} &	=  (\mathbf{r}_i - \mathbf{r}_1)^T (\mathbf{s}_j - \mathbf{s}_1). \label{con_4}
\end{eqnarray}
By this process there are $(m-1) (n-1)$ equations of type (\ref{con_4}) which are bilinear in $\rrr$ and $\sss$, which has rank�$3$ as discussed before. 

By factorizing $\hat{\mathbf{B}} = \tilde{\mathbf{R}}^T \tilde{\mathbf{S}}$, we can almost solve the node calibration problem, however there are a few remaining unknowns:
\begin{itemize}
\item Since the factorization is not unique. If $\hat{\mathbf{B}} = \tilde{\mathbf{R}}^T \tilde{\mathbf{S}}$, then also
$\hat{\mathbf{B}} = \tilde{\mathbf{R}}^T A A^{-1} \tilde{\mathbf{S}}$ is a valid factorization.
\item The starting sender positions $\sss_1$ is unknown.
\item The starting receiver positions $\rrr_1$ is unknown.
\end{itemize}

However, since the choice of coordinate system is arbitrary anyway, one may without loss of generality set the origin to $\rrr_1$. Also since any matrix $A$ can be QR-factorized as a rotation matrix times a triangular matrix, one may assume that $A$ is triangular, i.e.\ $A=L$. Thus fixing the rotation of the coordinate system. The remaining unknowns can be parametrized as
\begin{equation} \label{eq::risj}
	\begin{split}
		& \mathbf{r}_1=\mathbf{0}, \\
		& \mathbf{s}_1=\mathbf{Lb }, \\ 
		& \mathbf{r}_i =\mathbf{L}^{-T}\tilde{\mathbf{R}}_i, \ i=2 \ldots m, \\ 
		& \mathbf{s}_j =\mathbf{L} ( \tilde{\mathbf{S}}_j + \mathbf{b} ), \ j=2 \ldots n,
	\end{split}
\end{equation}
where $\tilde{\mathbf{R}} = \mathbf{L}^{T} \mathbf{R}$, $\tilde{\mathbf{S}} = \mathbf{L}^{-1} \mathbf{S}$, and hence $\hat{\mathbf{B}} = \tilde{\mathbf{R}}^T \mathbf{L}^{-1} {\mathbf{L}} \tilde{\mathbf{S}}= {\mathbf{R}}^T  {\mathbf{S}}$.

Using this parametrization, the equations (\ref{con_1}, \ref{con_2}, \ref{con_3}) become
%\begin{eqnarray}\label{parameter}
%	&d_{11}^2& 							= (\mathbf{r}_1 - \mathbf{s}_1) ^T (\mathbf{r}_1 - \mathbf{s}_1)  = \mathbf{s}_1^T \mathbf{s} _1\nonumber 
%	= \mathbf{b}^T \mathbf{L}^T \mathbf{L} \mathbf{b} \label {con_11} \\
%	&&=  \mathbf{b}^T \mathbf{H}^{-1} \mathbf{b}, \label {con_111} \\
%	&d_{1j}^2 - d_{11}^2 &	=  \mathbf{s}_{j}^T\mathbf{s}_{j} - \mathbf{s}_{1} ^T\mathbf{s}_{1}  \nonumber
%	= \tilde{\mathbf{S}}_j ^{T} \mathbf{L} ^T \mathbf{L} \tilde{\mathbf{S}}_j + 2 \mathbf{b} ^T\mathbf{L}^T \mathbf{L} \tilde{\mathbf{S}}_j \label{con_22} \\
%	&&= \tilde{\mathbf{S}}_j ^{T} \mathbf{H}^{-1} \tilde{\mathbf{S}}_j+ 2 \mathbf{b}^T \mathbf{H}^{-1} \tilde{\mathbf{S}}_j,\label{con_222} \\
%	&d_{i1}^2 - d_{11}^2 &	=  \mathbf{r}_{i}^T\mathbf{r}_{i} - 2 \mathbf{r}_{i}^T\mathbf{s}_{1} \nonumber
%	= \tilde{\mathbf{R}}_i ^{T} (\mathbf{L} ^T \mathbf{L})^{-1} \tilde{\mathbf{R}}_i- 2 \mathbf{b}^T  \tilde{\mathbf{R}}_i  \label{con_33} \\
%	&&=  \tilde{\mathbf{R}}_i ^{T}  \mathbf{H} \tilde{\mathbf{R}}_i- 2 \mathbf{b}^T\tilde{\mathbf{R}}_i, \label{con_333}
%\end{eqnarray}
\begin{eqnarray}\label{parameter}
	&d_{11}^2& =  \mathbf{b}^T \mathbf{H}^{-1} \mathbf{b}, \label {con_111} \\
	&d_{1j}^2 - d_{11}^2 &	=  \tilde{\mathbf{S}}_j ^{T} \mathbf{H}^{-1} \tilde{\mathbf{S}}_j+ 2 \mathbf{b}^T \mathbf{H}^{-1} \tilde{\mathbf{S}}_j,\label{con_222} \\
	&d_{i1}^2 - d_{11}^2 &	=    \tilde{\mathbf{R}}_i ^{T}  \mathbf{H} \tilde{\mathbf{R}}_i- 2 \mathbf{b}^T\tilde{\mathbf{R}}_i, \label{con_333}
\end{eqnarray}
where we have introduced the symmetric matrix $\mathbf{H}=(\mathbf{L} ^ T \mathbf{L})^{-1}$.

If enough correspondences are given then we can use the equations of type (\ref{con_333}) to estimate $\mathbf{H}$ and $\mathbf{b}$ linearly. 
Then we can backtrack and calculate $(\mathbf{r}_1, \ldots, \mathbf{r}_m, \mathbf{s}_1, \ldots, \mathbf{r}_m)$ using equations (\ref{eq::risj}). 

If there are both fewer than $10$ sound events and fewer than $10$ microphones, we can still solve the problem, but we have to use the non-linear equations (\ref{con_111}) and (\ref{con_222}). This special case can be solved using polynomial equation solving as described in \cite{kuang-burgess-etal-icassp13}.

The initial estimates are then refined by minimizing the non-linear least squares problem,
\begin{eqnarray}
\min_{\mathbf{m}_i,\mathbf{s}_j,o_j} \sum_{ij} \left( u_{ij} - ( || \mathbf{m}_i - \mathbf{s}_j||_2 + o_j ) \right)^2 
\end{eqnarray}
using standard techniques (Levenberg-Marquart) in order to obtain the maximal likelihood estimate of the parameters. 

%See Algorithm~\ref{alg:alg_3}.
%%--------------------------------------------------------------------------------------
%\IncMargin{1em}
%\begin{algorithm}
%\SetKwData{Left}{left}\SetKwData{This}{this}\SetKwData{Up}{up}
%\SetKwFunction{Union}{Union}\SetKwFunction{FindCompress}{FindCompress}
%\SetKwInOut{Input}{input}\SetKwInOut{Output}{output}
%\Input{Raw matches $\mathbf{u}^A$, offsets $\mathbf{o}^B$}
%\Output{C-matches $\mathbf{u}^C$, C-results ($\mathbf{x}^C, \mathbf{y}^C, \mathbf{o}^C)$}
%\BlankLine
%\emph{{\bf{Set parameters:}} $\mathbf{x}$ for microphone positions, $\mathbf{y}$ for sound positions and $\mathbf{o}$ for offsets.}
%\vskip 3mm
%\emph{1. Remove offsets that would cause $\mathbf{d} = \mathbf{u} - \mathbf{o} $ to be negative.\\
%2. Check for errors.\\
%3. Recalculate $\mathbf{d}$ with proper offsets.\\
%4. Calculate $\mathbf{x}$ and $\mathbf{y}$ using minimal solver by \cite{kuang-burgess-etal-icassp13}.\\
%5. Check for errors.\\
%5. Remove outliers after bundling.}
%
%\caption{Estimate microphone, sound position and refine offset (Step C)}\label{alg:alg_3}
%\end{algorithm}\DecMargin{1em}

\section{Find more inliers from time-difference estimation}
\label{sec:retrack}

For find more inliers among the remaining columns in $u_{ij}$, first we calculate microphone positions $\mathbf{r}_i$ and offsets $o_j$ by trilateration using time differences vector $\mathbf{u}$ and sound position $\bf s$. For each time instances in inlier set of $\mathbf{u}$, trilaterate one point $\mathbf{r}_0$ using RANSAC followed by bundle adjustment, which is non-linear least square optimization of $\mathbf{m_i}$ with $\mathbf{m_0}$ as initial estimate. 
\begin{eqnarray}
\min_{\mathbf{m_i}} \sum_{j} \left( u_{ij} - ( || \mathbf{m_i} - \mathbf{s}_j||_2 ) \right) .
\end{eqnarray}

From $I$ and $J$ which is indeces of inlier set that $I$ represents indeces in channels, 1 to $M$, and $J$ represents indeces from $1$ to number of matches construct vector $\bf D$, it takes the values in $u$ which is in inlier set. Then calculate the residuals by $res = \sqrt{[{\bf m}(J) - {\bf s}(I)]^2} + o(J) - \bf D$.
We set when $res < 0.05$, we count number of  inliers in each column if it is $\ge 5$, we set it as inliers.

\section{Robust estimation of mirrored microphones}
\label{sec:mirror}

The output for the system so far is typically the microphone positions corresponding to the direct path, i.e.�$ \rrr_{1,1}, \ldots, \rrr_{m,1}$ and sound event positions $\sss_j$ at a number of different times $t_j$. 
Lets return to equation (\ref{eq:uikik}), 
\begin{equation}
w_{i_1,k_1,i_2,k_2} =  || \sss - \rrr_{i_2,k_2} || - || \sss - \rrr_{i_1,k_1} || . 
\end{equation}

In this section we present a robust method for finding mirrored sound source positions $\rrr_{i,k}$. 
From the previous system we have already established some direct path microphone position $\rrr_{i,1}$ and
some sound event positions $\sss$. Lets study the the time-difference measurements from microphone positions to
multipath microphone position $\rrr_{i,k}$:
\begin{equation}
\begin{split}
w_{1,1,i,k} =  & || \sss - \rrr_{i,k} || - || \sss - \rrr_{1,1} || , \\
w_{2,1,i,k} =  & || \sss - \rrr_{i,k} || - || \sss - \rrr_{2,1} || , \\
& \vdots \\
w_{m,1,i,k} =  & || \sss - \rrr_{i,k} || - || \sss - \rrr_{m,1} || . \\
\end{split}
\end{equation}
By rearrangement
\begin{equation}
\label{eq:ChannelConsistency}
\begin{split}
w_{1,1,i,k} + || \sss - \rrr_{1,1} || =  & || \sss - \rrr_{i,k} || , \\
w_{2,1,i,k} + || \sss - \rrr_{2,1} || =  & || \sss - \rrr_{i,k} || , \\
& \vdots \\
w_{m,1,i,k} + || \sss - \rrr_{m,1} || =  & || \sss - \rrr_{i,k} || . \\
\end{split}
\end{equation}
we see that we should find the same shape for all of the $m$ different tdoa-images. 
In Figure~\ref{fig::ChannelConsistency} we show the top 4 peaks of the GCC-PHAT signal for each time instant corrected for direct path distance according to equation (\ref{eq:ChannelConsistency}). Notice that some curves are identical in the peaks. The cross channel consistent data is shown to the right. 

\begin{figure}
\centering
\def\svgwidth{5cm}
\includegraphics[height=2.5cm,width=9cm]{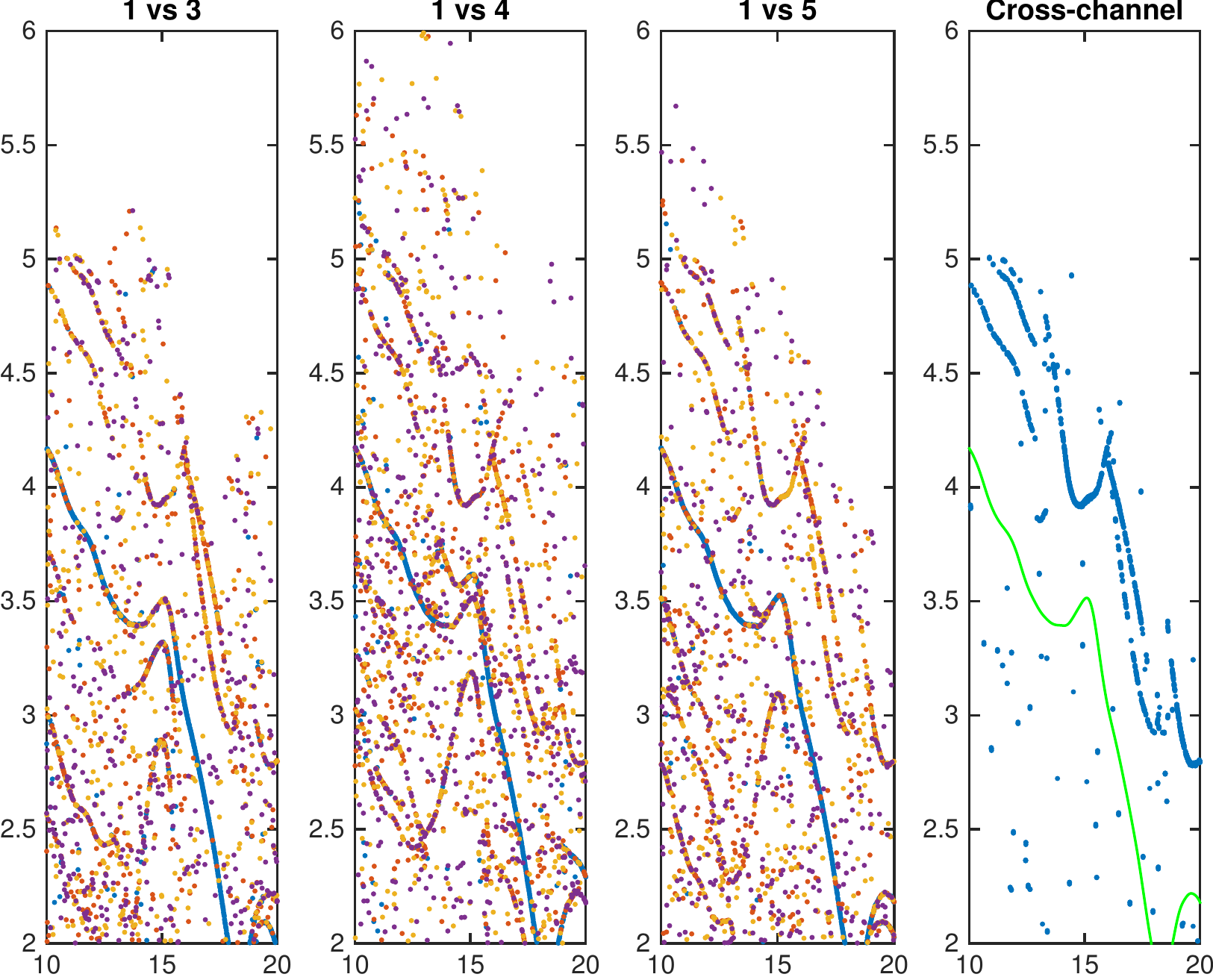}
\caption{The figure shows GCC-PHAT peaks for three channels corrected for direct path distances according to equation (\ref{eq:ChannelConsistency}) for channel pairs (1 vs 3), (1 vs 4) and (1 vs 5). To the right is shown those data that are consistent over all seven channel pairs (1 vs 2), \ldots (1 vs 8). TDOA measurements that are consistent over these channels are candidates for multipath components. }
\label{fig::ChannelConsistency}
\end{figure}

Using the cross channel consisten data (Figure~\ref{fig::ChannelConsistency}) we get candidates for multipath component
distances  $|| \sss - \rrr_{i,k} ||$. Again since we already have an estimate on sound source paths, the problem has only three degrees of freedom in, i.e.\ the components of $\rrr_{i,k}$. Again we use a hypothesize and test paradigm. 

\begin{enumerate}
\item Choose three random points among the cross channel consistent data.
\item Find the two candidate positions for the multi-path position of the microphone. 
\item For each candidate position $\rrr$ calculate $ || \sss - \rrr ||�$ and check for inliers. 
\item Repeat steps 1-3. 
\item Choose the point $\rrr$ with the most inliers. 
\item Perform local non-linear optimization to minimize sum of squared reprojection errors among the inliers.
\end{enumerate}

After finding one mirrored position remove the inlier data and repeat the process. 

\begin{figure}
\centering
\def\svgwidth{5cm}
\begin{tabular}{cc}
\includegraphics[height=2.5cm,width=4cm]{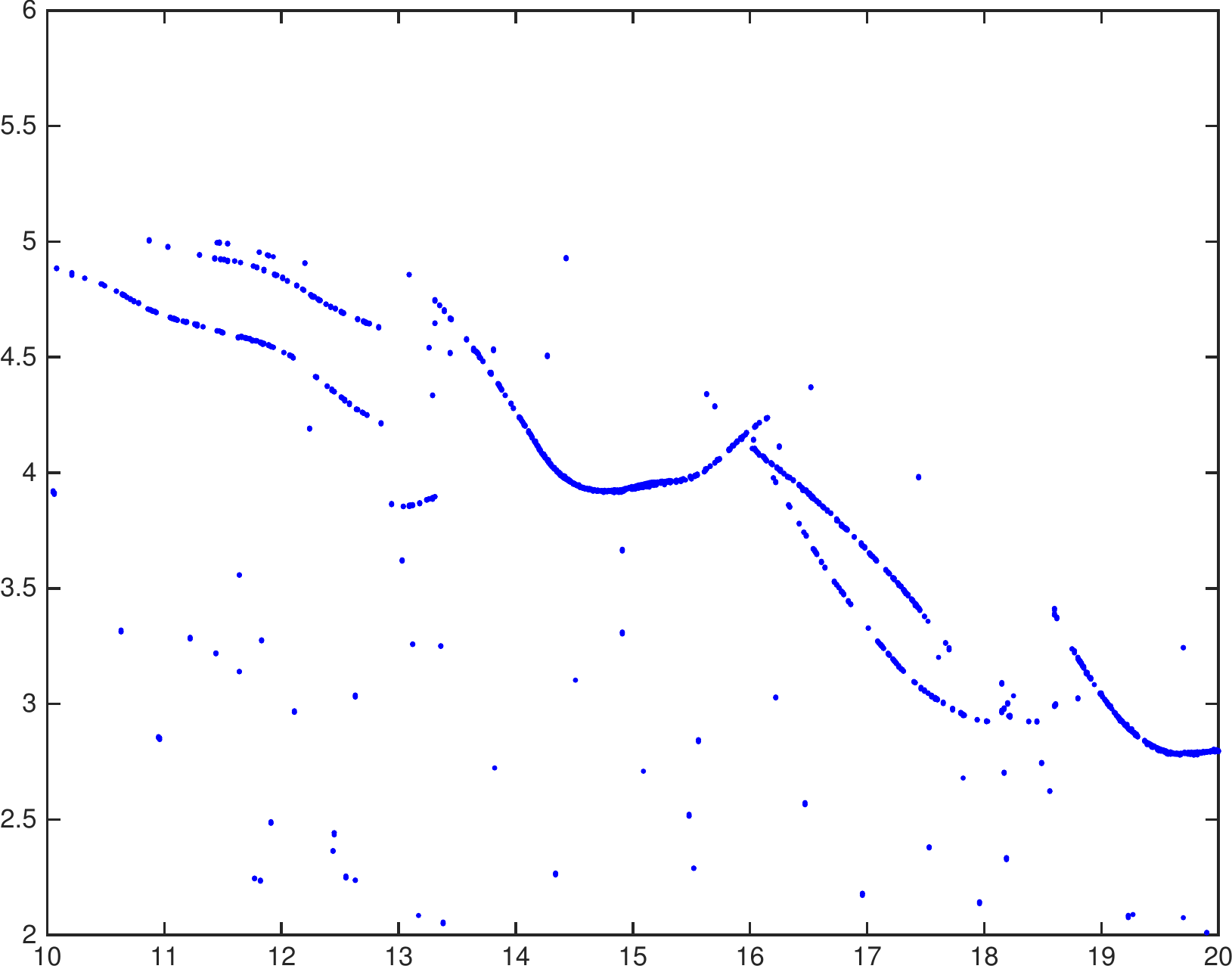} & 
\includegraphics[height=2.5cm,width=4cm]{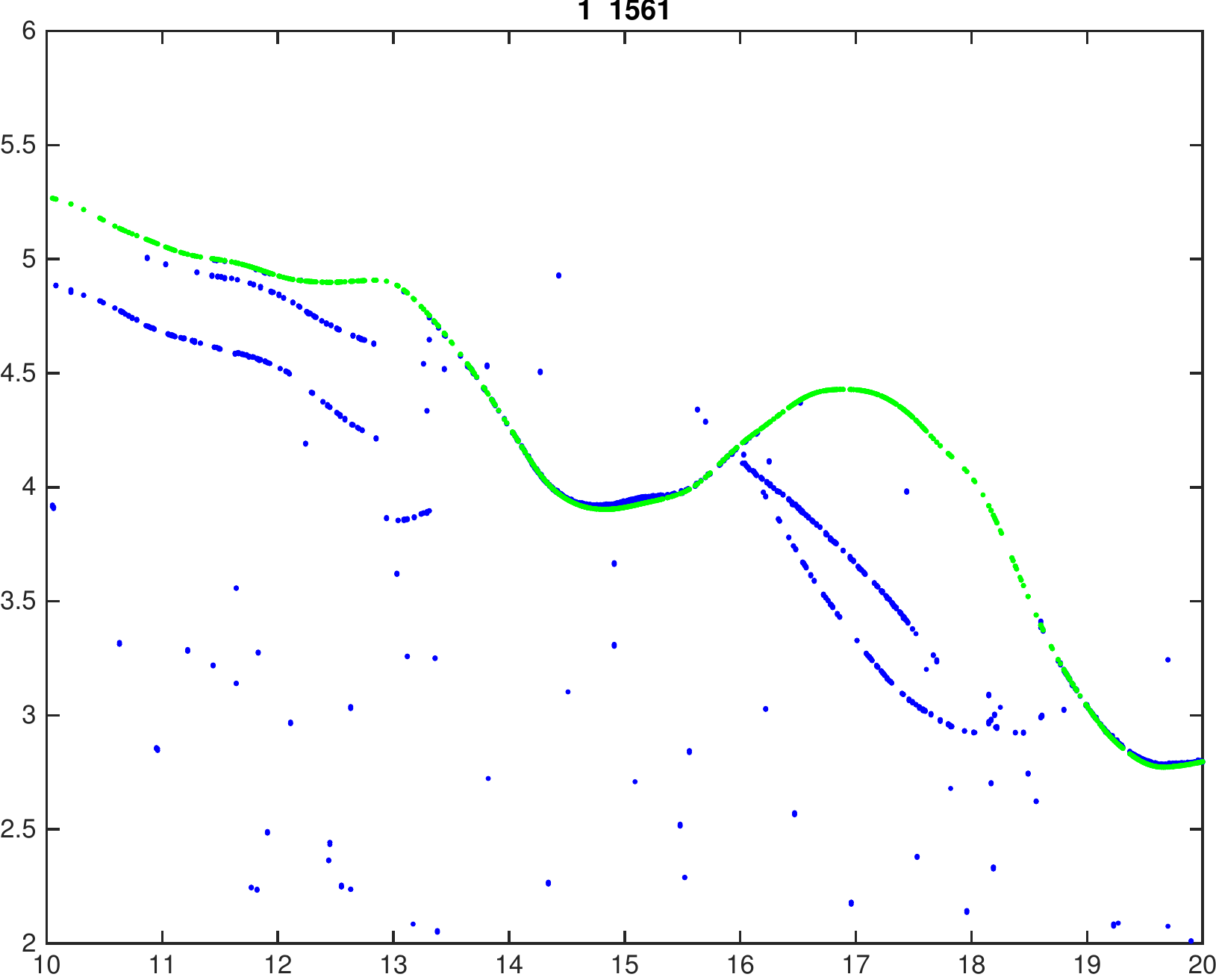} \\
\includegraphics[height=2.5cm,width=4cm]{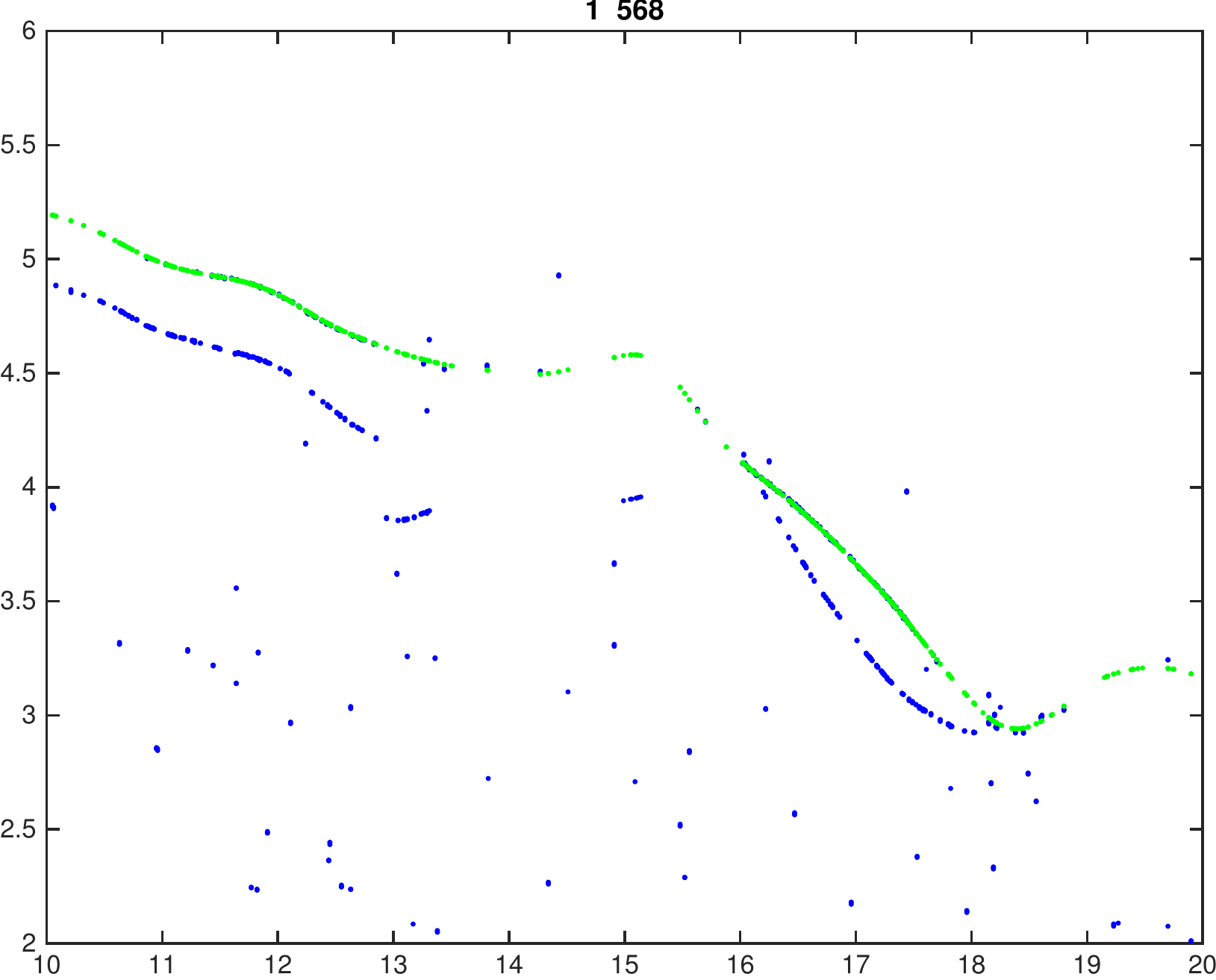} &
\includegraphics[height=2.5cm,width=4cm]{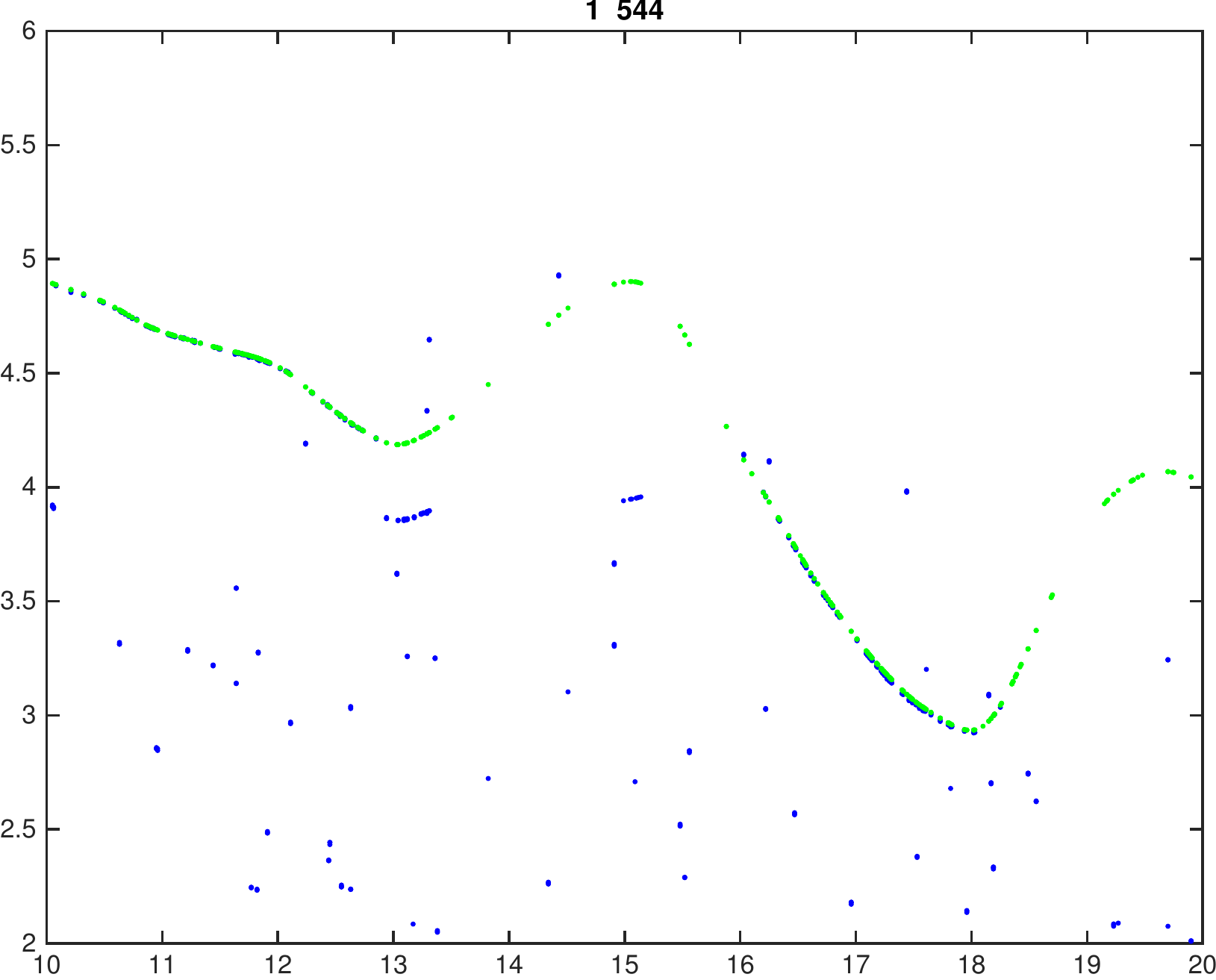} \\
\end{tabular}
\caption{The figure shows the candidates for multipath components (top-left). After running RANSAC once we obtain a candidate for a multi-path microphone position consistent with 1561 inliers (top-right). After running RANSAC once more on the remaining data we obtain another candidate for a multi-path microphone position consistent with 564 inliers (bottom-left). After running RANSAC once more on the remaining data we obtain another candidate for a multi-path microphone position consistent with 544 inliers (bottom-right).}
\label{fig::ChannelConsistency}
\end{figure}

This process is then repeated for the other channels so as to find multi-path mirrored microphone positions for all microphones. 

\section{Experimental Validation}
\label{sec:experiment}

We have made several experiments to verify the different components of the system as well as the robustness and performance of the whole system. 

We first study the numerical stability of the minimal solvers (Sections~\ref{sec:o}-\ref{sec:rs}).
Then we study the performance of the tracking algorithms of Section~\ref{sec:tracking}.
Then follows evaluations of the whole system for a couple of different scenarios.

%We have made several experiments as described in Section~\ref{}.
%with 8 microphones (Shure SV100). These are connected to an audio interface (M-Audio Fast Track Ultra 8R) connected to a laptop. The microphones were positioned in a room with approximate distance \mbox{0 - 2} meters from each other.  
%We generated sounds in several scenarios which is:
%\begin{itemize}
%\item Random distinct sound bursts made by banging two spoons together. This produces a set of discrete sound events that are relatively easy to detect and match. 
%\item One continuously moving sound source playing part of a song. This produces a set of smoothly changing time-differences. If this is known, tracking techniques (Kalman filter, Particle filter) could be used to track the changes. 
%\item Several continuously moving sound sources.
%\item Mixture of several people talking, clapping, walking around in the room. 
%\end{itemize}
%The 8 sound channels were sampled at 96000 Hz. 

\subsection{Synthetic Data}\label{sec:numstab}

\begin{figure}
\center
\begin{tabular}{c c}
\includegraphics[width= 0.48 \columnwidth]{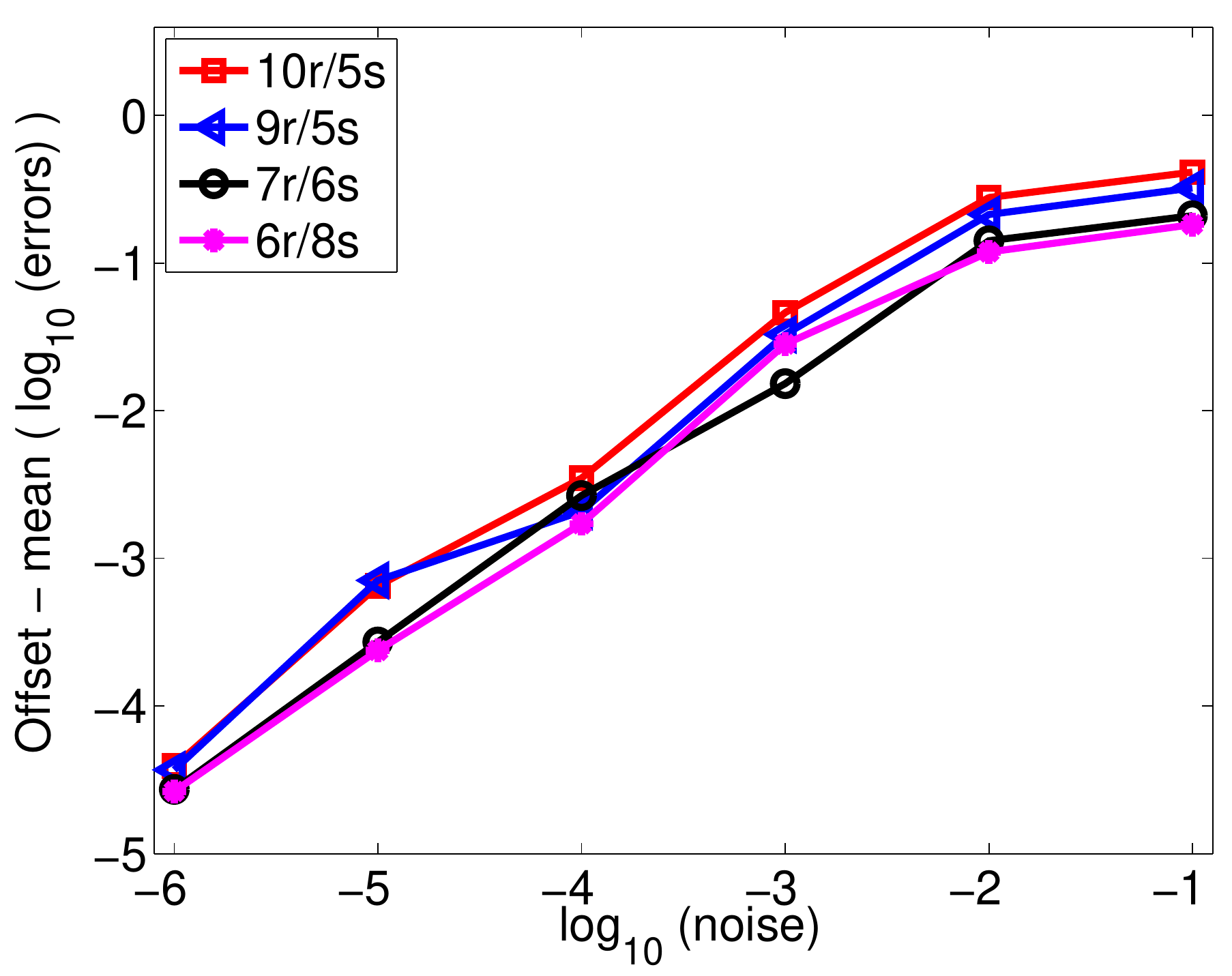}
\includegraphics[width= 0.48 \columnwidth]{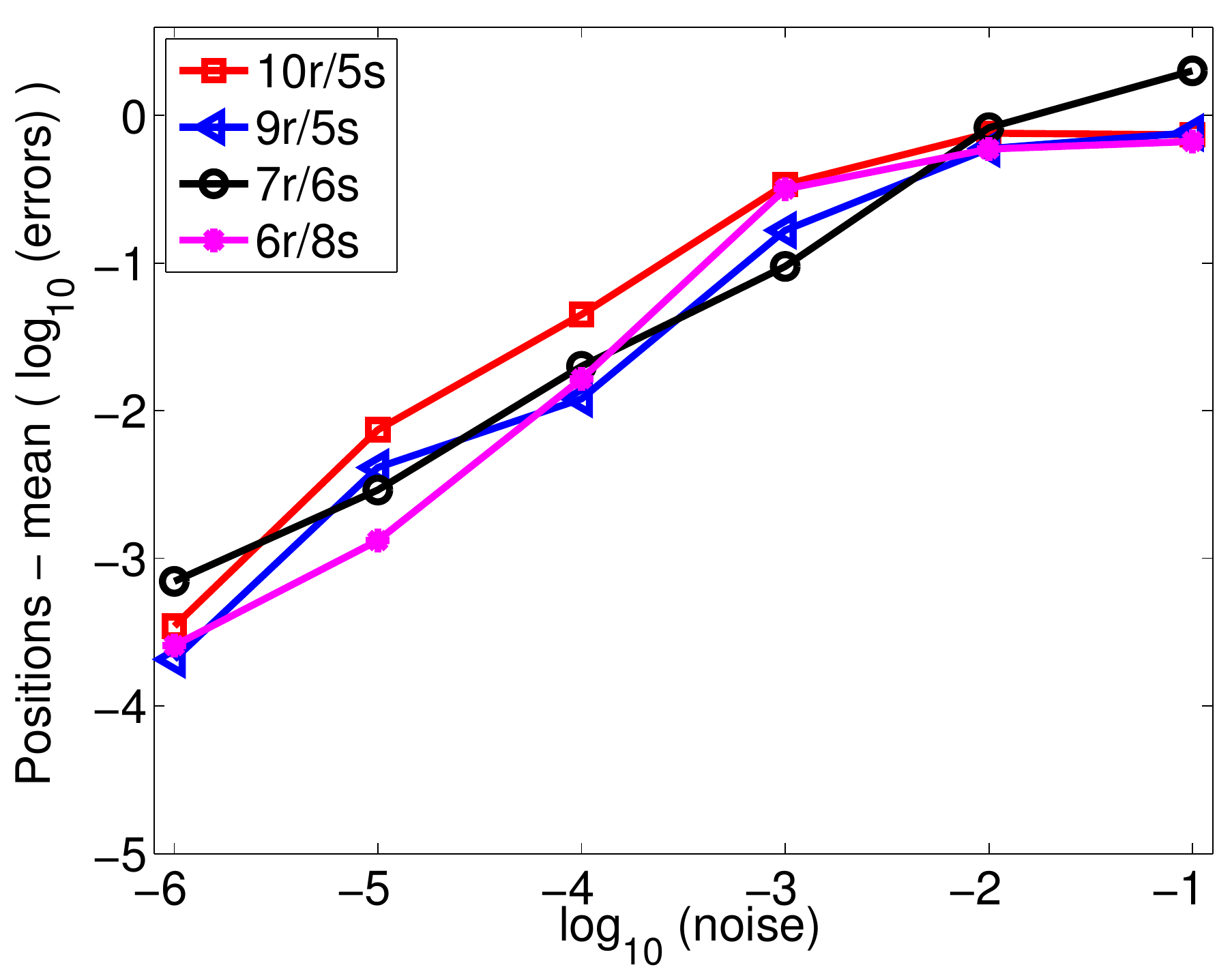}  \\
\includegraphics[width= 0.48 \columnwidth]{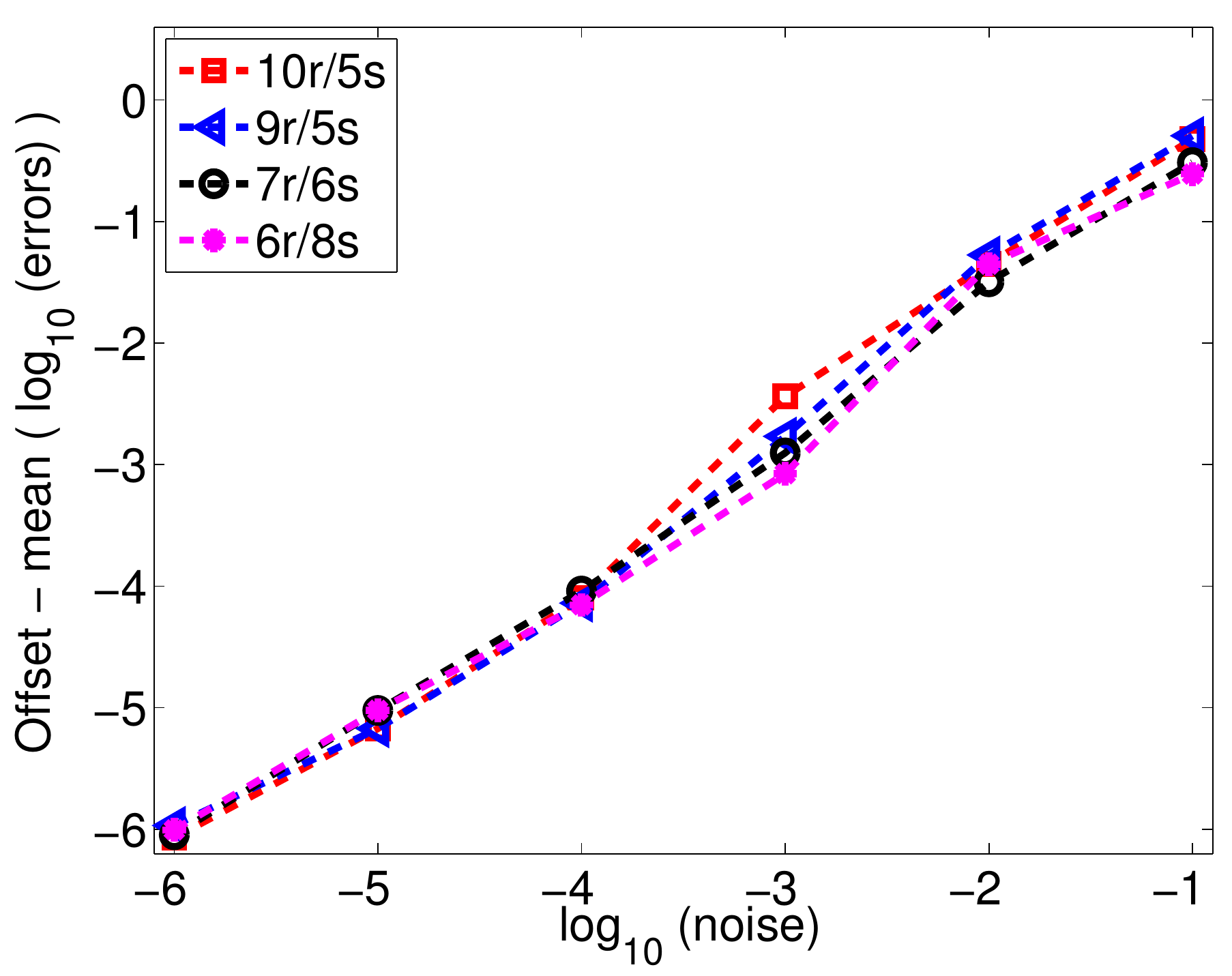}
\includegraphics[width= 0.48 \columnwidth]{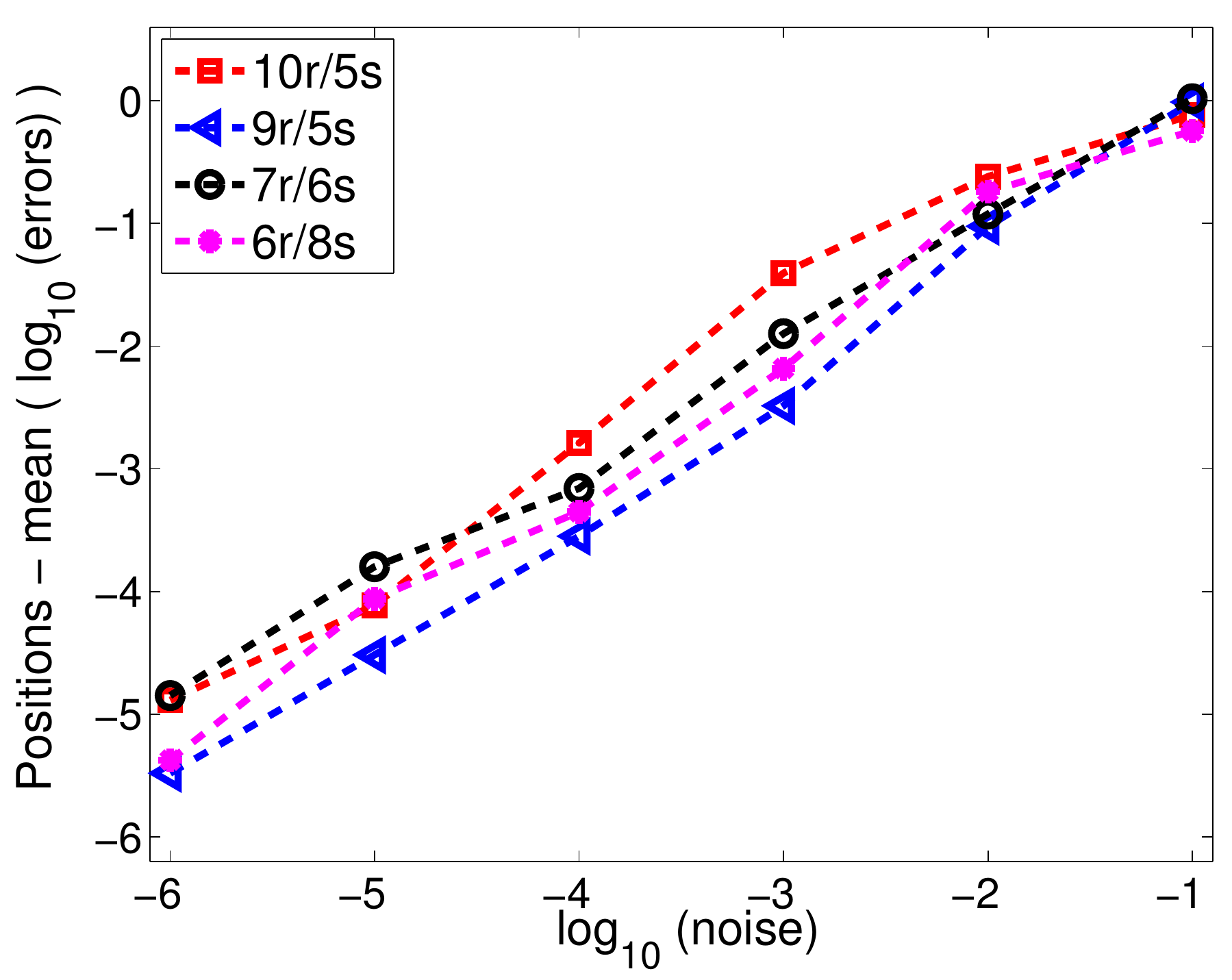}
\end{tabular}
\caption{Synthetic experiments for TDOA solvers on 3D under Gaussian noise. The errors of estimated time offsets (left) and reconstructed positions of microphones and sounds (right) are shown. (Top) Performance of different solvers ($10$r$/5$s \cite{pollefeys-nister-icassp-08}, $9$r$/5$s, $7$r$/6$s and $6$r$/8$s) with their corresponding minimal settings for solving offsets; (Bottom)  with  20 receivers and 20 microphones.}
\label{fig_synthetic}
\end{figure}

\begin{figure}
\center
\begin{tabular}{c c}
\includegraphics[width= 0.50 \columnwidth]{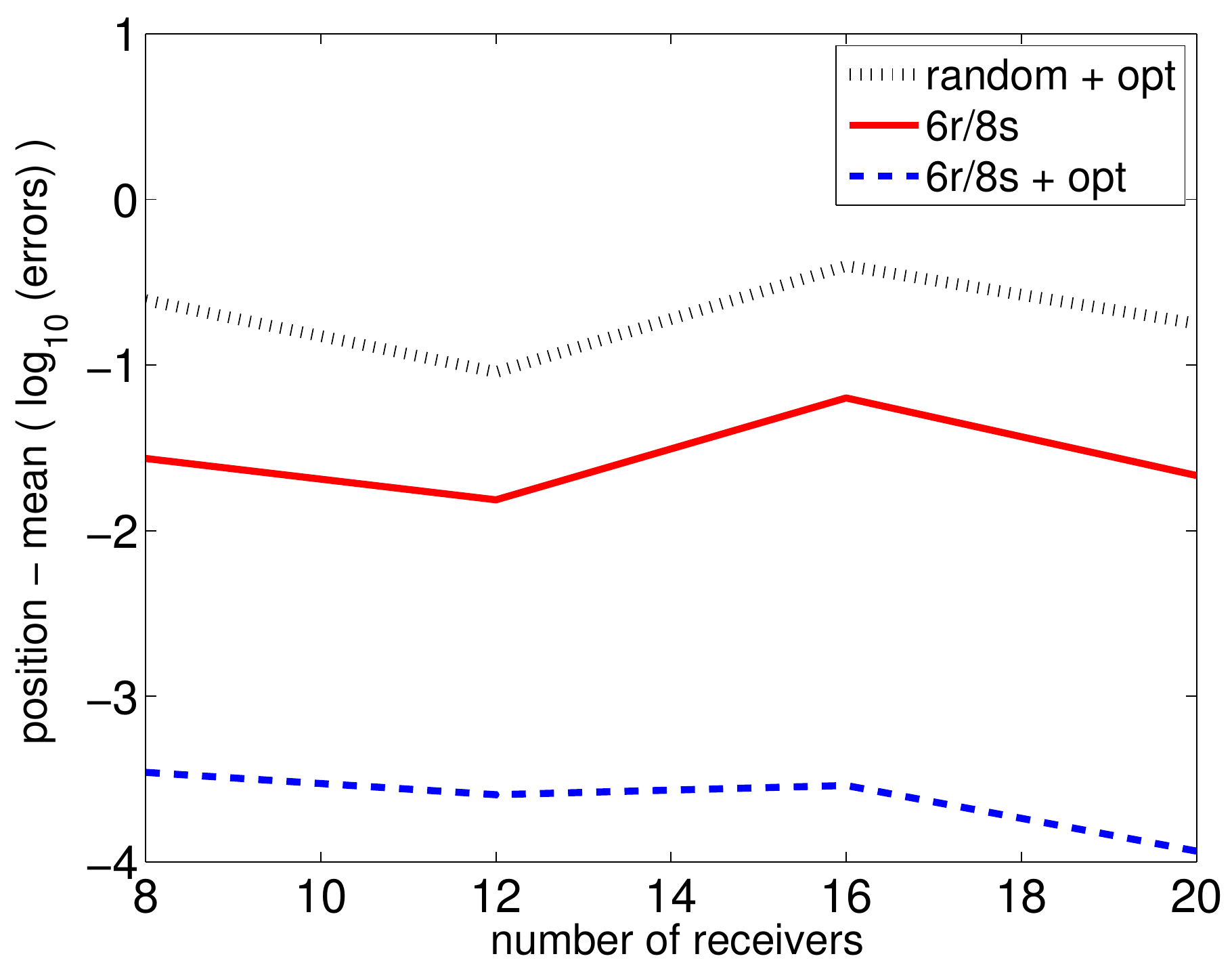}
\end{tabular}
\caption{Initialization with random offsets and our $6$r$/8$s solver with varying number receivers (8 to 20) and 8 transmitters (noise level $10^{-4}$) for non-linear optimization. }
\label{fig_compa}
\end{figure}

In this section, we study the numerical behaviors of the TDOA solvers on synthetic data. 
We simulate the positions of microphones and sounds as 3D points with independent 
Gaussian distribution of zero mean and identity covariance matrix. As for the offsets, 
we choose them from independent Gaussian distribution with zero mean and standard 
deviation $10$.  We study the effects of zero-mean Gaussian noise on the solvers, 
where we vary the standard deviation of the Gaussian noise added to the TDOA 
measurements. When solving TOA problem, we have used the scheme discussed 
in Section \ref{sec:rs} for over-determined cases. To compare the reconstructed 
positions of microphones and sounds with the true positions, we rotate and translate 
the coordinate system accordingly. We can see that from Fig. \ref{fig_synthetic}, 
our proposed solvers $9r/5s$, $7r/6s$ and $6r/8s$ give numerically similar results 
as the $10r/5s$ case in \cite{pollefeys-nister-icassp-08} for both minimal settings 
and over-determined cases. In Fig.\ref{fig_compa}, random initialization of the time offsets resulted in poor convergence in the non-linear optimization, while our method provides with a much better starting point. On the other hand, we have also compared our solver with the iterative method proposed in \cite{jiang-etal-icassp13} for estimating the offsets. The method in \cite{jiang-etal-icassp13} converges very slowly ($5 sec.$ - $1 min.$ on a standard laptop, especially for (near) minimal settings) and tends to converge to the wrong local minima. While our solvers perform consistently well for all cases, they are also much faster (approximately $0.5s$ for the unoptimized codes). 
%This suggests the usability of our proposed solvers 
%in RANSAC as well as for practical settings with 
% with limited availability of the receivers. 

\subsection{Claps}

For this experiment we used $8$ Shure SV100 microphones recorded at 44.1kHz in an office. 
They are approximately 0.3-1.5 meters away from each other and are placed so that they span 3D.
%The placements of the microphones are not on a plane. 
We connected them to an audio interface (M-Audio Fast Track Ultra 8R), which is connected to a computer. 
%The 8 synchronized channels are . 
We generated sounds by moving around in the room and clapping approximately 1-2 meters from the microphones. We collected 5 independent recordings of approximately 20s. Each recording contained roughly 30 claps (sound events).  

\begin{figure}
\vspace{-6mm}
\center
\begin{tabular}{c}
\includegraphics[width= 0.5\columnwidth]{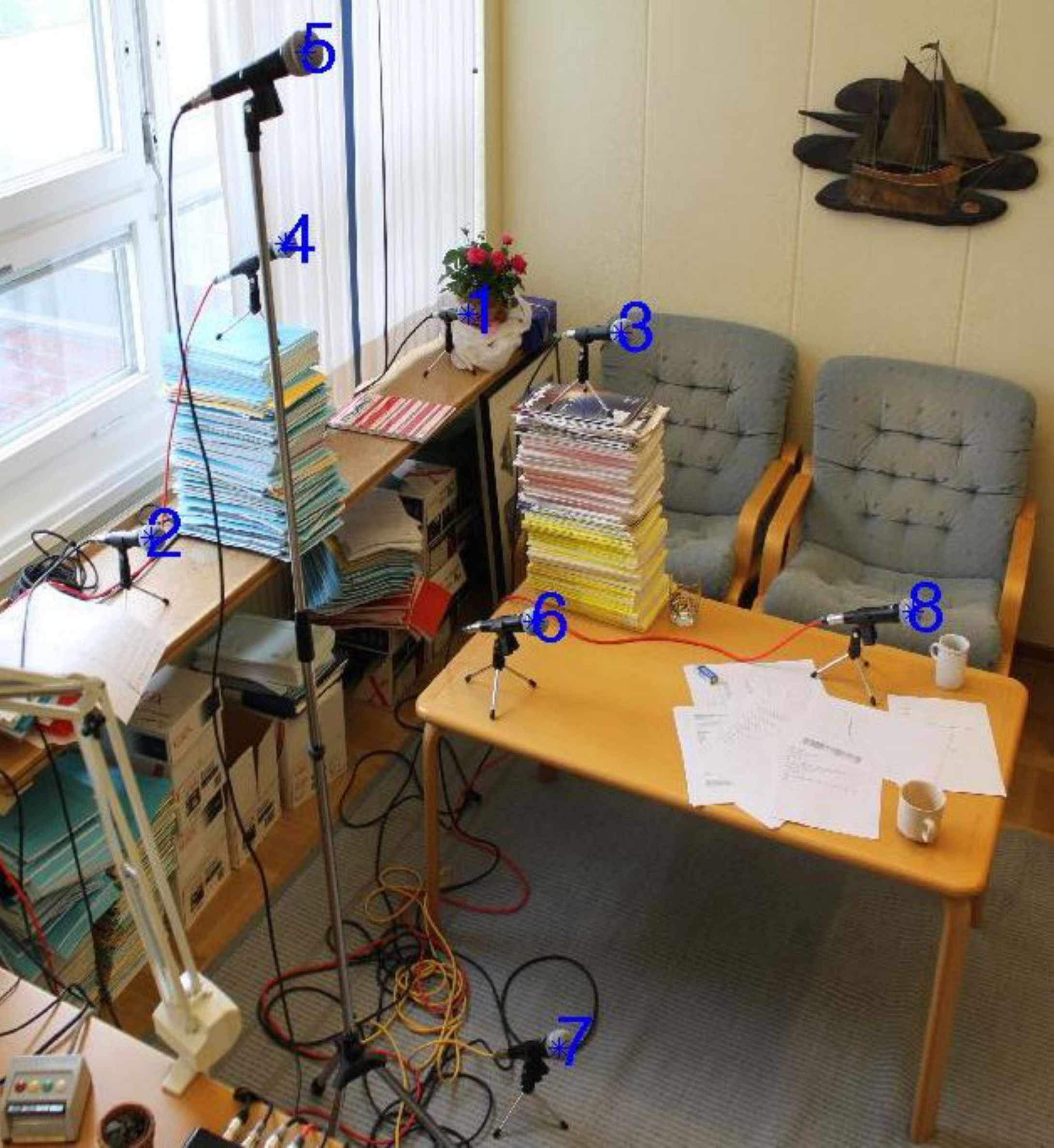}\\
\includegraphics[width= 0.63\columnwidth]{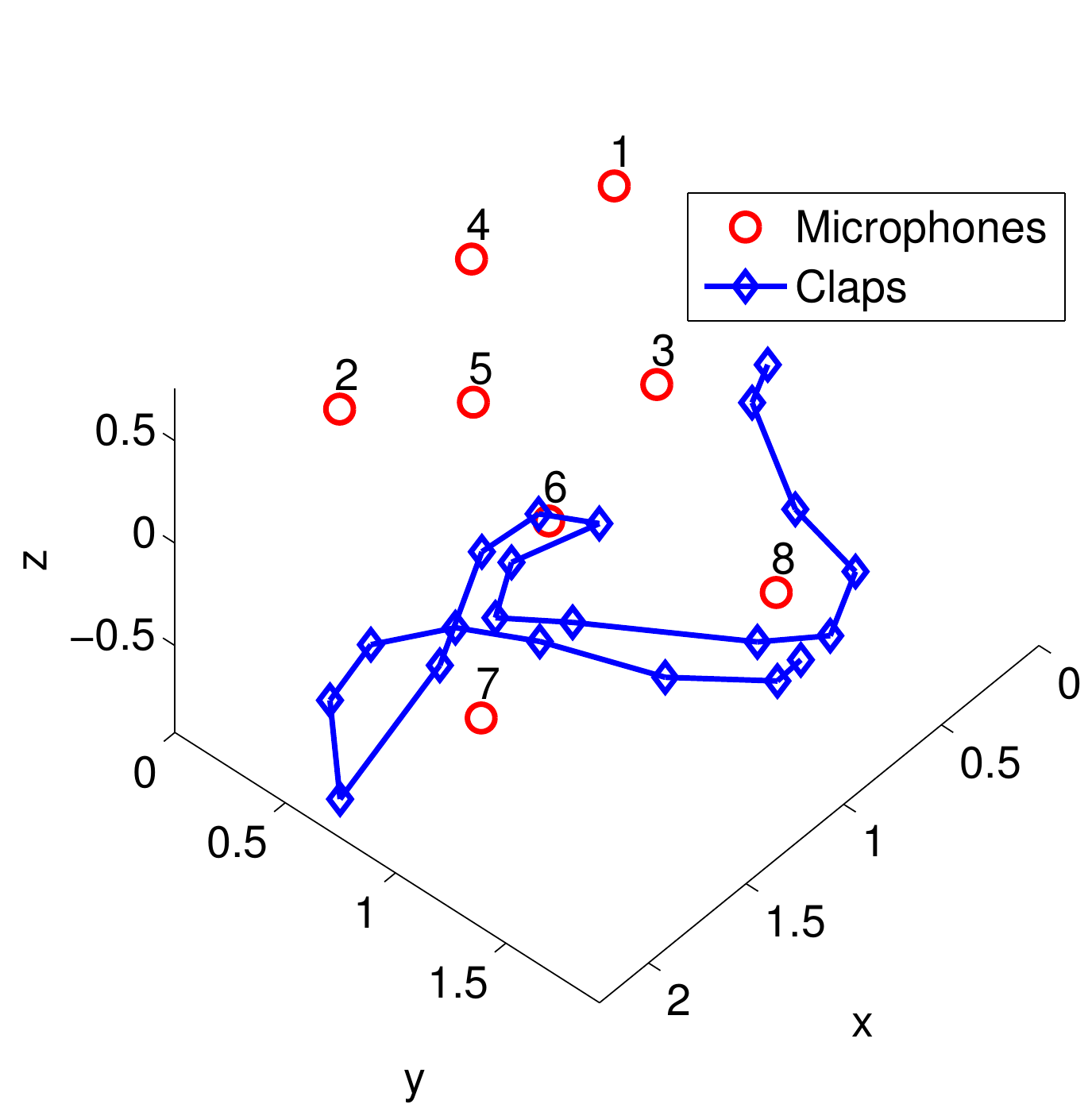} \\
\includegraphics[width= 0.63\columnwidth]{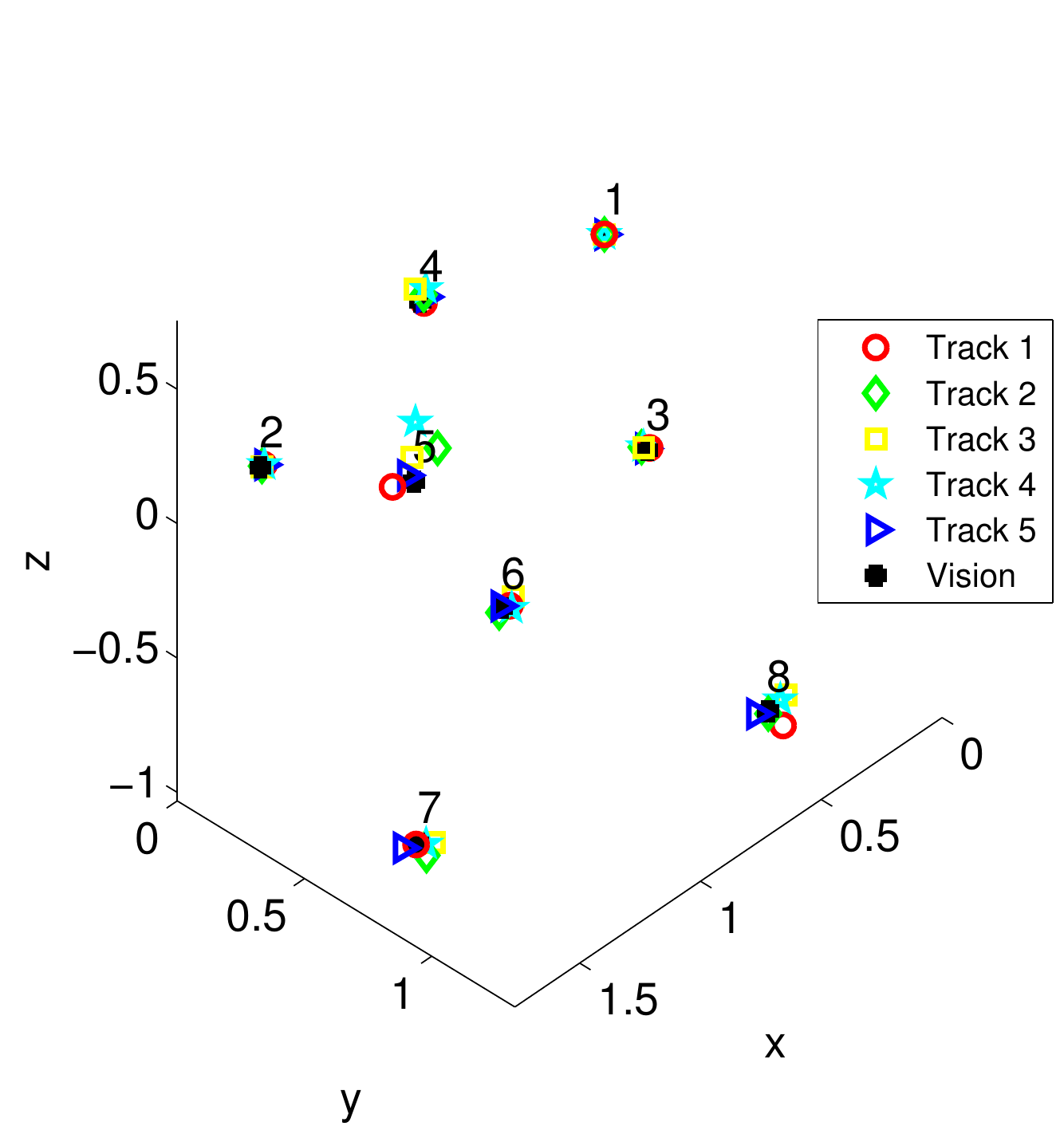}  \\
\end{tabular}
\caption{Results on TDOA with microphones and sounds.  Top: View of the experimental setup. Middle : Reconstruction of 8 microphones (in red - 'o') and 21 sound events (blue - '$\diamond$') from one of 5 independent recordings; Right : Reconstructed microphone positions estimated from 5 independent experiments and compared to reconstruction using computer vision (black - '+') }
\label{fig_real}
\end{figure}

To obtain TDOA measurements, we coarsely matched sounds of the claps to sound flanks as described in Section~\ref{sec:tracking}.A.
%The matching scheme is coarse and  can be improved with generalized cross-correlation.
%have defined interest points on the recording as edges between periods with low energy and periods with high energy. 
For the experiment we used only those claps that were detected in all 8 channels. 
We ran both the $7$r$/6$s and $6$r$/8$s solvers to determine the offsets followed by an alternating optimization that refines the offset estimation. After solving the unknown transformation and translation, we recover an initial euclidean reconstruction for the locations of microphones and claps. Finally we refine the reconstruction with non-linear optimization. The result of one of these 5 reconstructions are shown in Figure~\ref{fig_real} (middle). The reconstructed microphone positions from 
these 5 independent multi-channel recordings were put in a common coordinate system and compared
to each other. The average distance from each microphone to the its corresponding mean position (estimated from corresponding reconstruction of the 5 recordings) is 2.60 cm. It is important to point out that without proper initialization using our methods, the solutions we get converge poorly (with large reconstruction errors). Previous solvers do not work here due to either insufficient number of receivers (10 receivers needed in \cite{pollefeys-nister-icassp-08}) or violating the assumption that one of the microphones collocates with one of the claps \cite{crocco-delbue-etal-tsp-2012} .

As an additional evaluation, we have also reconstructed the locations of the microphones based on computer vision techniques. We took 11 images of the experimental setup. 
Figure \ref{fig_real} ({left}) shows one of the 11 images used. 
We manually detected the 8 microphone center positions in these 11 images and used 
standard structure from motion algorithms to estimate the positions of the
8 microphones. 
%Here we assumed unit aspect ratio, zero skew, principal point in the middle of the image and estimated one common focal length as well as calibrated structure from motion. 
The resulting reconstruction is also compared to that of the five structure from sound reconstructions. 
The comparison is shown in Figure~\ref{fig_real} ({right}). 
We can see the TDOA-based reconstructions are consistent with the vision-based reconstruction.

\subsection{Continuous songs in anechoic chambers}

We have made several experiments using a continuously moving sound source in anechoic chambers. These produced GCC-PHAT data that were relatively easy for our system to track and match over time. Even in the case of two sound sources moving simultaneously and continuously in the chamber produced relatively good results. 

We illustrate some of the steps of the automatic system with one of the experiments. In this case we have two moving sound sources in 3D. We assume $c = 343~m/s$ in room tempreture. 
%Figure~\ref{fig::matchingscore} shows a plot of the matching score for different shifts between the channel $4$ and channel 1, i.e. $u_4 = c (t_4 - t_1)$ on the y-axis at $999$ equally spaced positions along the $48$ seconds long recording. 
The matching algorithm produces $129$ matching vectors. There are $83$ missing data among these $129 \times 8 = 1032$ time difference measurements. The RANSAC algorithm finds an inlier set of $75$ (out of the $129$) matching vectors. %This is illustrated in Figure~\ref{fig::ransac} where each dot corresponds to a measurement. Missing data are indicated as absence of a dot. The RANSAC algorithms selects random subsets of $7$ rows and $6$ columns. One such random selection is illustrated with blue dots. The inlier data from the algorithm are illustrated with green dots. 

\begin{figure}
\centering
\def\svgwidth{5cm}
\includegraphics[height=5cm,width=8cm,trim = 0mm 70mm 0mm 70mm]{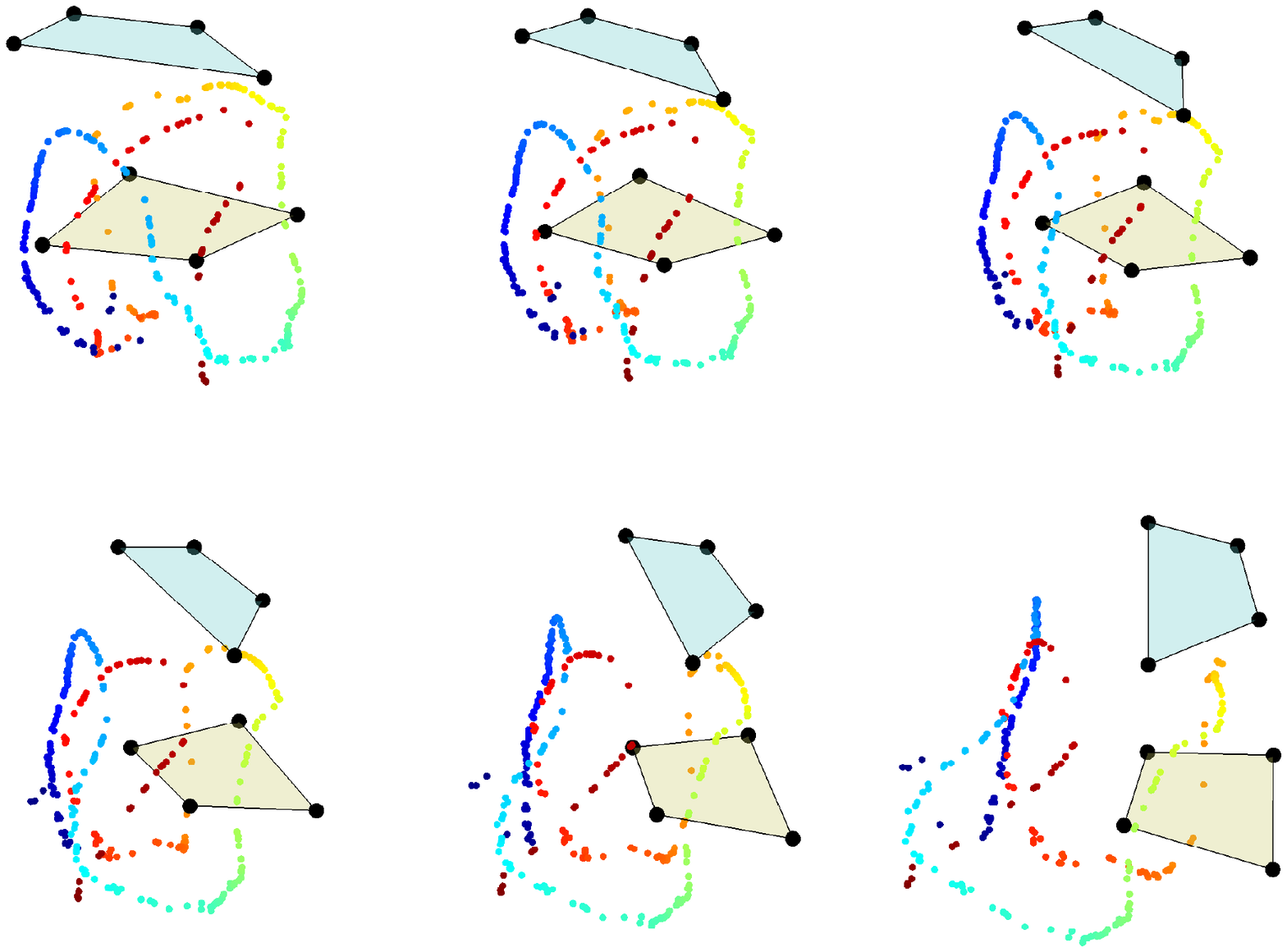}
\caption{Three-dimensional reconstruction of the microphone setup (black dots) as well as three-dimensional reconstruction of moving sound source positions (coloured dots). Note that the microphones in the experiment setup are located in two planes (also indicated in the figure).}
\label{fig::3dplot}
\end{figure}

\begin{figure}
\centering
\def\svgwidth{5cm}
\includegraphics[height=4cm,width=8cm]{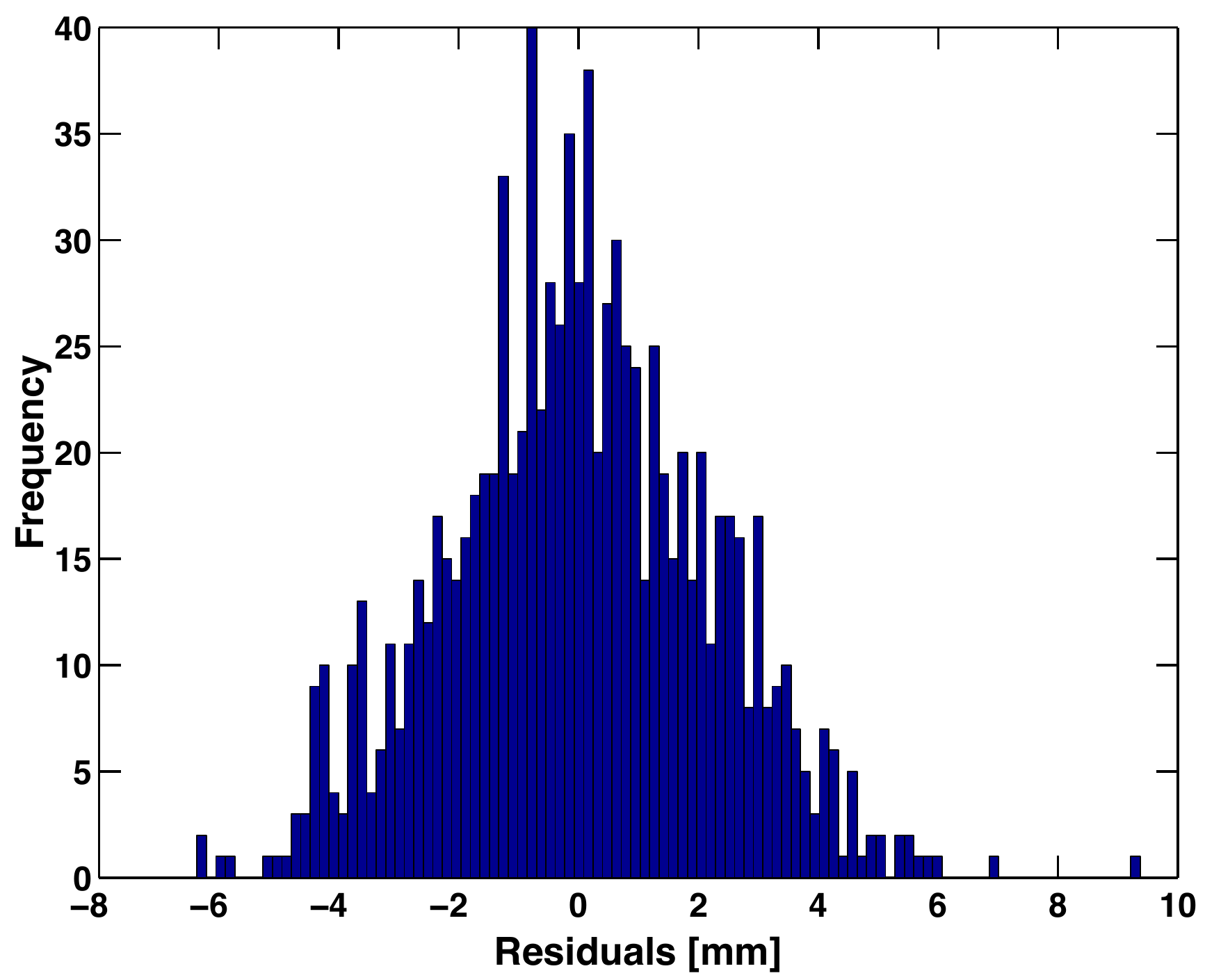}
\caption{Histogram of the residuals between the measured data $u_{ij}$ and the fit $|| \mathbf{m}_i - \mathbf{s}_j||_2 + o_j$.}
\label{fig::residuals}
\end{figure}

These 75 inlier matching vectors are then used to estimate the 3D positions for the senders and receivers. A histogram of the residuals $u_{ij} - ( || \mathbf{m}_i - \mathbf{s}_j||_2 + o_j ) $ is shown in
Figure~ \ref{fig::residuals}. The errors are in the order of a few millimeters. The final 3D reconstruction of the microphones and of the sound source paths for one of the experiments are shown in Figures~\ref{fig::3dplot}. In this experiment we have microphones in two planes (four in each). The moving sound source starts outside the convex hull of the microphones, then moves inside the microphone cluster and then out again. 

After refinement step, we get 129 matches.

To validate the method we have used several independent recordings. These have different sound source positions, but identical microphone setup. The error between the three reconstructions and the mean has a standard deviation of about 1 cm. %shown in Figure~ \ref{fig::ComparisonofReconstruction}, indicating the accuracy of the system. 

%\begin{figure}
%\centering
%\def\svgwidth{5cm}
%\includegraphics[height=6cm,width=8cm]{ComparisonofReconstruction.eps}
%\caption{Reconstructions of microphone position using three different data set (with same microphone position).}
%\label{fig::ComparisonofReconstruction}
%\end{figure}

\subsection{Matching in reverberant surroundings}
\label{sec:falsepositives}

Several experiments were carried out. During these experiments we used eight T-bone MM-1 microphones, connected to a M-Audio Fast Track Ultra 8R sound card and recorded with the program Audacity on a MacBook Pro with 3 GHz Intel Core i7 processor. Recordings were made in different rooms (seminar rooms, lunch rooms etc). The experimental environments were chosen as to have a few planar surfaces (floor, walls, ceilings) with relatively low sound absorption coefficient as well as a few other objects such as chairs, tables, flower pots etc. Several experiments were made with a single moving sound source, i.e.\
a Roxcore Portasound loudspeaker connected to a mobile phone. In the experiment we used different songs.

For the experiment we generated ground truth matching difference vectors for the direct path. After each of the steps of the matching system, as described in Section~\ref{sec:tracking}, we computed the detections obtained by the automatic system with the ground truth and calculated the number of inlier detections and outlier detections. These are shown in Table~\ref{tab:res}. Notice that the number of inliers are kept at a high level, while the number of outliers decrease. The output of the system is essentially outlier free.

%Resultat:
%* Hur minskar FPR? Hur minskar/ökar TPR?
%* Hur gör systemet med gamla trackern och med nya.
%* Hur ser residualerna
%* Procrustes på mikrofonpositioner
%Jag lyckades inte fa bassh3- att fungera i det gamla systemet. Har for mig att jag har fatt det att fungera en gang, kan handa att man inte kan anvanda samma settings...

\begin{table}[thb]
  \small
  \begin{center}
    \renewcommand{\arraystretch}{1.2}
    \begin{tabular}{lccc}
      \hline
      \textbf{Step} & Inliers & Outliers & Corresponding figure\\
      \hline
      Step 1& 1260 & 4639 & \\%Figure~\ref{fig:toppeaks}\\
      Step 2& 1228 &  633 & Figure~\ref{fig:matchedpeaks}\\
      Step 3& 1027 &   30 & Figure~\ref{fig:ransac}\\
      Step 4& 1199 &   93 & \\ %Figure~\ref{fig:segments}\\
      Step 5& 1199 &    7 & \\%Figure~\ref{fig:connectedsegments}\\
      Step 6& 1475 &   17 & Figure~\ref{fig:smooth}\\
      Step 7& 1275 &    0 & \\%Figure~\ref{fig:smooth}\\
      \hline
    \end{tabular}
  \end{center}
%  \vspace*{-0.3cm}
  \caption{The number of outliers and inliers after each step of the algorithm.}
  \label{tab:res}
\end{table}

%\begin{figure}[thb]
%  %\centering
%  \setlength\figureheight{0.4\linewidth}
%  \setlength\figurewidth{0.8\linewidth}
%  \tikzsetnextfilename{residuals}
%  \input{figures/residuals.tikz}
%  \vspace*{-6mm}
%  \caption{Residuals for one channel.}\label{fig:residuals}
%\end{figure}

%%%%%%%%%%%%%%%%%%%%%%%%%%%%%%%%%   temporary put a comment %%%%%%%%%%%%%%%%%%%%%%%%%%%%%%%%%
%\begin{figure}[thb]
%  %\centering
%  \setlength\figureheight{0.3\linewidth}
%  \setlength\figurewidth{0.6\linewidth}
%  \renewcommand{\titlefortikz}{Projected mirrored delays}
%  \tikzsetnextfilename{spegling12}
%  \input{figures/spegling12.tikz}
% % \vspace*{-4mm}
%  \caption{Different paths calculated form the estimated microphone and sound source positions using trilateration for the second longest segment (representing reflection in the floor).}\label{fig:spegling12}
%\end{figure}
%%%%%%%%%%%%%%%%%%%%%%%%%%%%%%%%%%%%%%%%%%%%%%%%%%%%%%%%%%%%%%%%%%%%%%%%

%\begin{figure}[htb]
%  \centering
%  \includemedia[%
%    width=0.9\columnwidth,%
%    activate=pagevisible,%
%    deactivate=pageinvisible,%
%    %noplaybutton,%
%    %3Dtoolbar,%
%    3Dviews=figures/path3D.vws%
%  ]{%
%    \setlength\figureheight{0.4\linewidth}%
%    \setlength\figurewidth{0.7\linewidth}%
%    \tikzsetnextfilename{path3D}%
%    \input{figures/path3D.tikz}%
%    %\includegraphics[width=0.5\textwidth]{figures/path3D}%
%  }{%
%    figures/path3D.u3d%
%  }
%  \vspace{-6mm}
%  \caption{Estimated path of the sound source microphone positions. (3D in Adobe Reader)}
%  \label{fig:path3D}
%\end{figure}

\subsection{System evaluation}

The detections of the system is then used as input to the estimation of the parameters $\rrr_i$, $\sss_j$ and $o_j$
as described in Sections~\ref{sec:o}-\ref{sec:rs}. For visualization purposes we have calculated the camera matrix $P$ for the image in relation to the world coordinate system obtained by the audio reconstruction. This makes it possible to visualize the estimated motion path of the sound source in the video. We also made several independent recordings with the microphones in the same positions. The reconstructed microphone positions could then be compared with each other, thus estimating the reconstruction errors. The estimated standard deviation in microphone coordinates was $3.7$ millimeters.

\begin{figure}[thb]
  \centering
  \setlength\figureheight{0.5\linewidth}
  \setlength\figurewidth{0.9\linewidth}
  \renewcommand{\titlefortikz}{Mirrored microphones}
  \tikzsetnextfilename{speglingar}
  \input{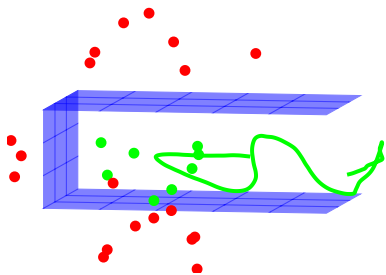}
%  \vspace*{-6mm}
  \caption{The figure illustrates the estimated microphone positions (green dots), estimated mirrored microphone positions (red dots) and sound source motion (green curve). The estimated reflective planes are also shown in the figure. These three planes correspond to the floor, the ceiling and the walls as seen in Figure~\ref{fig:illustration}}\label{fig:speglingar}
\end{figure}

\section{Conclusion}
\label{sec:conclusions}

In this paper, we have developed a novel automatic microphone self-localization using ambient sound. The system does not put any constraints on the motion of the sound sources in relation to the microphone array setup. The system is based on a first finding several time-difference matching vectors for the recording. These are then used as input to robust geometric algorithms based on minimal solvers and RANSAC to provide initial estimates of the unknown parameters, i.e.\ the offsets and the 3D positions of the sound sources and the receivers. For this purpose we have developed efficient algorithms that solve for offsets using minimal data. There are three such minimal problems assuming that microphones and sound sources span 3D. There are two such minimal problems assuming that either the microphones or the sound sources are restricted to a plane. For the case of microphones or sound sources restricted to a line there is one such minimal problem. The robust estimation algorithm produces an initial estimate of the geometry parameters and a set of inlier measurements. These estimates of the geometry parameters are then improved by non-linear optimization using the inlier data to obtain the maximum likelihood estimate of the parameters. We have also used sub-sample refinement of the time-difference-of-arrival measurements and shown that they improve the precision of the system. The resulting microphone accuracy is in the order of millimeters. The components of the system as well as the system as a whole has been tested on both synthetic and real data with promising results. 

Although we have solved the minimal problems for offset estimation using a rank constraint, it is still an open problem to study and solve the minimal problems for solving the TDOA problem using all constraints. Other possible avenues for future research would be to improve on the overall system, making it even more robust. Here different strategies for tracking could be tried out on real and synthetic data. The code for the minimal solvers as well as the system is published on github, \url{http://github.com/kalleastrom/StructureFromSound}. We also provide on our homepage a set of example recordings that can be used with the system, \url{http://vision.maths.lth.se/sfsdb/}. We hope that this can be used to improve on the system.

\bibliographystyle{plain}
\bibliography{tdoa_toa}

\end{document}